\documentstyle[aps,prc,multicol,epsfig]{revtex}

\begin{document}

\title{Shrinking $\hbar$ as a recipe for revealing classical-like
contributions\\ in optical potential cross sections}

\author{R.Anni}

\address{
Dipartimento di Fisica dell'Universit\`a, Universit\`a di Lecce, Lecce,
Italy\\
Istituto Nazionale di Fisica Nucleare, Sezione di Lecce, Lecce, Italy\\
}

\maketitle

\begin{abstract}
% DON'T CHANGE THIS LINE
%
A simple recipe for revealing classical-like contributions in optical
potential cross sections is proposed. 
The recipe is based on the fact that the classical-like properties are
not expected to depend on the actual value of $\hbar$. 
This allows us to identify the classical-like characteristics of an
optical potential cross section by simply repeating the calculation with
different values of $\hbar$, and observing which properties of the cross
section are invariant.
This method is applied to the cross sections of a few optical
potentials used to describe the recent data of light heavy-ion elastic
scattering.
\end{abstract}


\begin{multicols}{2}

\section{Introduction}

The recent detailed measurement of the elastic differential cross
sections of light heavy-ions\cite{KHO00,NIC00,NIC99,OGL00} have
stimulated a revival of the interest in heavy-ion elastic scattering as
a useful tool to obtain a better understanding of nuclear properties.

Deep real parts and shallow imaginary ones characterise the optical
potentials used to reproduce these experimental cross sections.
The reduced absorption, that seems necessary to reproduce the observed
data, makes the cross section sensitive to contributions coming from
the internal region of the interaction, allowing the study of the
interaction properties in regions where the densities of the two nuclei
strongly overlap.

The number of partial waves that contribute to these optical potential
cross sections is in all these cases rather large.
This allows one to hope that semiclassical techniques can be used to
connect the behavior of the cross sections, in certain angular ranges,
to the properties of the interaction, in some corresponding spatial
region.

The recognition of the presence of classical-like contributions in an
experimental cross section may be a very difficult job, but at least in
principle, their determination in an optical potential cross section
should be easier.
Here the problem is in fact reduced to performing a semiclassical
analysis of the optical potential cross section.

The classical limit of scattering by a real potential is well
understood since more then fourty years\cite{FOR59}.
This limit is found with the use of asymptotic approximations for the
scattering function $S(\lambda)$ ($\lambda=l+\frac {1}{2}$), for the
Legendre polynomials $P_l(\cos \theta)$, and for the partial wave
expansion of the scattering amplitude $f(\theta)$.
As a result one obtains that $f(\theta)$ can be expressed in terms of
one, or more, stationary phase contributions.
The square modulus of each stationary phase contribution exactly
coincides with a corresponding contribution from one branch of the 
classical deflection function.
Because each stationary point contribution has also a phase, if two or
more stationary points contribute to $f(\theta)$ in some angular range,
oscillations appear in the cross section arising from interference.
These oscillations disappear in the classical limit by averaging over
the finite resolution of the experimental devices.

Unfortunately this simple scheme cannot, in general, be applied to the
optical potentials commonly used to describe the elastic scattering of
two heavy-ions.
In the scattering process from the optical potentials currently used
(also neglecting the complications arising from the presence in the
interaction of an imaginary part) the classical integral actions, in
units of $\hbar$, are large but not infinite and quantum contributions,
superimposed to classical-like ones, survive in the scattering function
and in the corresponding cross section.
Only in the extreme classical limit these quantum contributions are
expected to disappear, remaining confined in regions whose width go to
zero.

Several semiclassical methods\cite{BRI85} were developed to extend the
range of application of the classical description to the dynamical
regions in which an imaginary part is present in the potential and the
extreme classical conditions are not fully satisfied.

These methods considerably extend the possibility of describing the
semiclassical properties of a scattering process.
However, presently, they predict cross sections in good quantitative
agreement with the exactly calculated ones only in certain energy
ranges, depending on the optical potential considered.

Furthermore, also admitting that we are within the range of validity of
some of the available semiclassical methods, their application to
practical cases is slightly more difficult, and much less popular, than
the direct calculation of the exact cross section.
Owing to this, one could ask if some trick exists, which is able to
provide quickly a useful indication on the classical-like nature of
some properties of a cross section, without worrying about the
complications of the semiclassical techniques and about their ranges of
validity.

In this paper one of these possible tricks is investigated. 
The base idea is that the classical-like properties must not depend on
$\hbar$.
Owing to this all the properties of a cross section which do not depend
on the value which is attributed to $\hbar$, in the framework of a
quantum calculation, can be considered of classical origin.

The main ingredients of this simple recipe, which can be easlily
implemented in any standard optical potential code, are presented in
Sect. II.
In order to test the method in a simple case, in Sect. III we present
the results obtained for a real optical potential.
In Sect. IV and V the method is applied to analyze the behavior of two
optical potential cross sections, fitted to the experimental data of
$^{16}$O +$^{12}$C at the laboratory energies of 132 and 200
MeV\cite{OGL00}.

For the cases considered, the results of the quantum calculation are
first compared with the corresponding classical ones.
This comparison is not really necessary for the application of the
recipe and is introduced here only to show the reliability of the
method to identify correctly classical-like properties.

The results obtained are that the qualitative behavior of the optical
potential cross sections smoothly changes with varying $\hbar$.
Considering the oscillations which appear in the cross section as
arising from the interference between simpler amplitudes, one observes
that with decreasing values of $\hbar$ some of these amplitudes
continue to modify their behavior, contributing to angular ranges of
decreasing width, while others become insensitive on any further
decrease of $\hbar$.
The former reveal their quantum origin, the latter their classical
nature.

The comparison of the behavior of the real cross section (calculated
with the true value of $\hbar$) with that of the fictitious cross
sections (calculated attributing to $\hbar$ values sufficiently small)
allows one to obtain easy indications on the classical-like properties
of the true cross section.

\section{Main ingredients of the recipe}

Accordingly to classical mechanics, the cross section for scattering
from a potential $V(r)$ is completely determined by the deflection
function

\begin{equation}
\label{ClaDef}
\Theta(\lambda)=\pi-2 \int_{r_0}^{\infty}
\frac{\lambda}{k r^2 \sqrt{1-\frac{V(r)}{E}-\frac{\lambda^2}{k^2 r^2}}}
dr,
\end{equation}
where $L=\lambda \hbar$ and $p=k\hbar$\footnotemark are, respectively,
the angular and linear momenta, and $r_0$ is the turning point which
delimitates from below the classically accessible region of the radial
motion.

\footnotetext{The appearance of $\hbar$ does not imply the use of quantum mechanics,
but only the choice of convenient units.}

The deflection angle $\Theta(\lambda)$ is connected to the scattering
angle $\theta(\lambda)$ through the relation

\begin{equation}
\theta(\lambda)=\arccos [\cos \Theta(\lambda)].
\end{equation}

Remembering that $\Theta(\lambda) \rightarrow 0$ for $\lambda
\rightarrow \infty$, if $\Theta(\lambda)$ is a monotonous function and
$|\Theta(\lambda)|<\pi$, the differential cross section is given by

\begin{equation}
\label{ClaSez}
\sigma(\theta)= \frac{\lambda(\theta)}{k^2 \sin \theta}
\left|\frac {d \lambda(\theta)}{d \theta}\right|,
\end{equation}
where $\lambda(\theta)$ is the inverse function of $\theta(\lambda)$.

If the above conditions are not satisfied, particles with different
angular momenta can be scattered at the same scattering angle.
This happens when critical $\lambda$ values exist at which the
deflection function has maxima or minima, or crosses the values $-m
\pi$, $m=0,1,\ldots$.
In this case $\theta(\lambda)$ can be inverted only within intervals
limited by two consecutive critical $\lambda$ values, corresponding to
different branches of the deflection function, and the cross section is
given by

\begin{equation}
\label{MulClaSez}
\sigma(\theta)= \sum_j \frac{\lambda_j(\theta)}{k^2 \sin \theta}
\left|\frac {d \lambda_j(\theta)}{d \theta}\right|,
\end{equation}
where the sum runs over all the branches of the deflection function
containing a deflection angle corresponding to the scattering angle
$\theta$.

Accordingly to quantum mechanics, the cross section for scattering from
a potential $V(r)$ is completely determined by the scattering function
$S(\lambda)$.
This quantity defines the scattering amplitude

\begin{equation}
\label{QuaAmp}
f(\theta)=\frac{i}{k} \sum_{l=0}^{\infty}
\lambda [1-S(\lambda)] P_{\lambda-\frac{1}{2}}(\cos \theta),
\end{equation}
and the cross section $\sigma(\theta)$, which is the square modulus of
$f(\theta)$.

The classical limit of the quantum cross section is realized through
the appearance of a link between $S(\lambda)$ and $\Theta(\lambda)$.
These two quantities are in fact connected by the relation

\begin{equation}
\label{QuaLim}
\lim_{\hbar \rightarrow 0} \frac {d \arg S(\lambda)}{d \lambda}=\Theta(\lambda).
\end{equation}

Thanks to this link, the classical mechanics expression for the cross
section is recovered using the asymptotic expansion for the Legendre
polynomials, transforming the partial wave expansion into a sum of
integrals, and evaluating asymptotically these integrals using the
stationary phase method.

The properties of the deflection function fix the number of the
stationary phase points which contribute to $f(\theta)$ at each angle.
The contribution from each stationary phase point has a modulus, whose
square just coincides with the classical expression, and a phase.
The phase factors produce an oscillatory behavior in the cross section
in the angular regions where two, or more, stationary phase points
contribute, and the classical result is finally obtained only after
averaging over these oscillations, whose period goes to 0 in the
classical limit, to account for the finite resolution of the
experimental devices.

Owing to this, a signature of the contribution of classical-like
trajectories is the presence in the angular distribution of angular
intervals in which the cross section, changing the value attributed to
$\hbar$, either does not change or, if it changes, keeps as upper and
lower envelopes the curves corresponding to the maximal constructive
and destructive interference amongst all the contributions from the
different branches of the deflection function.
In the following these envelopes and the delimitated region will be
named {\it interference limits} and {\it interference region},
respectively.

The interference limits can be calculated starting from the properties
of $\Theta(\lambda)$, or they can be found by performing a quantum
mechanics calculation by attributing to $\hbar$ different values,
sufficiently smaller than the physical value.

This scheme is well suited to bring out classical like-contributions in
the cross section for scattering by a real potential, but it cannot be
directly applied to the cases in which an imaginary part is introduced
in the interaction to simulate the effects of the population of
channels different from the elastic one.
 
The imaginary  part $W(r)$ of the potential removes flux from the
elastic channel and the time dependence of the probability density,
$\rho({\bf r},t)=|\psi({\bf r},t)|^2$, of finding the scattering
partners at time $t$ at a relative position ${\bf r}$ satisfies the
equation\cite{BRI85a}

\begin{equation}
\frac{\partial \rho({\bf r},t)}{\partial t} + {\rm div} {\bf j}({\bf r},t)=
-\frac{2W(r)}{\hbar} \rho({\bf r},t),
\end{equation}  
where ${\bf j}({\bf r},t)$ is the probability current density. 
The above equation suggests the interpretation of the quantity
$w=2W(r)/\hbar$ as the probability per unit of time for a transition
out of the elastic channel.

With this interpretation of $w$  the form of the classical cross
section given by Eq. \ref{MulClaSez} remains the same, apart for the
introduction, in each term on the r.h.s, of a multiplicative factor

\begin{equation}
P(\lambda)=\exp \left( -\frac{1}{2\hbar}
\int_{r_0}^{\infty}\frac
{ \frac{W(r)}{E} }
{\sqrt{1-\frac{V(r)}{E}-\frac{\lambda^2}{k^2 r^2}}}
dr \right),
\end{equation}
expressing the probability that the particles with angular momentum
$\lambda \hbar$ are not removed from the elastic channel during their
motion along the classical trajectory.

We note, in passing, that this form for the cross section is just that
obtained using the naive WKB approximation\cite{BRO74} to estimate
$S(\lambda)$, and the stationary phase method to evaluate $f(\theta)$.
Within this scheme the probability that the particles with angular
momentum $\lambda \hbar$ are not removed from the elastic channel can
be identified with $|S(\lambda)|^2$, which has exactly the
dependence from the imaginary part of the potential deriving from the
above classical interpretation.

In accordance with this picture, in the search process of the
interference limits based on the variation of the value attributed to
$\hbar$ (in order to keep constant the survival probability factor) the
imaginary potential must be scaled with the same factor used for
$\hbar$.
This will be the additional caution used to look for traces of
classical-like contributions in the quantum mechanics cross section.

Apart from in the cross section, traces of classical-like contributions
can also be found in the behavior of $|S(\lambda)|$ and of $d \arg
S(\lambda)/ d \lambda$.
In the following this last quantity will be named {\it quantum
deflection function} and indicated with $\Theta_Q(\lambda)$.

If the classical limit is well approached both $|S(\lambda)|$ and
$\Theta_Q(\lambda)$ are expected not to depend on the value of $\hbar$,
if plotted versus the impact parameter $b=\lambda/k$.

The conventional optical potential codes provide the values of
$S(\lambda)$ for half integer positive $\lambda$ values.
These values can be used directly to plot $|S(\lambda)|$ versus the
impact parameter $b$, and using the finite difference formula

\begin{equation}
\label{QuaDef}
\Theta_Q(\lambda)=\frac {d \arg S(\lambda)}{d\lambda} \simeq 
\frac {\arg S(\lambda+\Delta\lambda) -\arg S(\lambda-\Delta\lambda)} 
{2 \Delta\lambda},
\end{equation}
with $\Delta\lambda=\frac{1}{2}$, they often also provide a reasonable
approximation for $\Theta_Q(\lambda)$ at integer $\lambda$ values.
In the cases here considered, for a very few $l$ values, this simple
approximation provides a bad estimate for $\Theta_Q(\lambda)$. 
This happens when the true variation of $\arg S(\lambda)$,
between consecutive integer $l$ values, becomes larger than 
$\pi$. 
It is due to the fact that, in the optical code that we use, the
arguments of $S(\lambda)$ are defined modulus $2\pi$ and the angle
$\alpha$ between two successive $S(\lambda)$ values, in the complex
Argand plane, is taken as the convex one.
Only for an extreme precaution, in the present work, the quantity
$\Theta_Q(\lambda)$ was estimated, outside our conventional optical
code, using Eq. \ref{QuaDef} with a step $\Delta \lambda=0.1$ for
$\alpha>\frac{\pi}{4}$ and of $0.5$ elsewhere.
 
The fact that the values of $|S(\lambda)|$ and $\Theta_Q(\lambda)$,
calculated with different values of $\hbar$ and plotted against $b$,
lie on the same curve can be considered a signature of the dominance of
a classical-like mechanism in the scattering process.
However this fact, alone, does not guarantee that the cross section
also closely follows the behavior predicted by the classical mechanism.
The realization of the classical limit for the cross section also
requires that the integrals, in which the partial wave expansion of
$f(\theta)$ is transformed, can be approximated by the stationary phase
method and that, at the stationary phase points, the Legendre functions
are well approximated by their non-uniform asymptotic expansions.

It is well known that the stationary phase method fails in a
neighbourhood of the classical rainbow angles and that the uniform
method allows an estimation of the contributions from the stationary
points in terms of Airy functions.
The uniform approximation substitutes the singularity of the classical
cross section, followed by the sharp shadow region, with a maximum in
the lit region followed by a decrease of the cross section.
In the approach to the classical limit the maximum moves towards the
rainbow angle, and the cross section very rapidly decreases in the
shadow region.
Therefore in general deviations are expected between the behavior of
the quantum and the classical cross sections around the classical
rainbow scattering angles.

Deviations are also expected in the extreme backward direction where
the usual non-uniform  approximation for the Legendre functions does
not hold.
This approximation is responsible for the presence of the factor
$1/\sin \theta$ in the classical limit of the cross section and,
consequently, for the appearance of the classical glory singularity.

These types of deviations, thypical of quantum effects, are however
standard and one can easily recognize their presence in the search
process of classical-like properties of the cross sections.

\section{Real optical potential cross section}

To test the effectiveness of the method based on the variation of
$\hbar$, we first consider the cross section of a fictitious real
optical potential having a conventional Woods-Saxon form factor with
parameters $V_{0}=282.2$ MeV, $R_{v}=2.818$ fm and $d_{v}=0.978$ fm.
This potential is the real part of one of those used for fitting the
elastic scattering cross section of $^{16}$O + $^{12}$C at
$E_{\rm Lab}=132$ MeV\cite{OGL00}.
For all the cases here considered, the Coulomb part of the interaction
is described using an analytical potential that closely approximates
the Coulomb potential of two uniformly charged spheres with radii of
3.54 fm and 3.17 fm.

\subsection{Comparison with the classical cross section}

The thick solid line in Fig. 1 shows the classical deflection  function
$\Theta(\lambda)$, as a function of the impact parameter $b$.
This line, as similar ones for the other cases considered, shows the
cubic spline interpolation of the $\Theta(\lambda)$ values calculated at
$b=\lambda/k$, with a step $\Delta\lambda=0.25$, starting from
$\lambda=0$.
The open dots in Fig. 1 show the values of $\Theta_Q(\lambda)$
estimated by using Eq. \ref{QuaDef} at integer $\lambda$ values, and
the thin curve gives the cubic spline interpolation of the calculated
points.

\begin{figure} 
\label{FIG1}
\hspace*{-3mm}
\epsfig{file=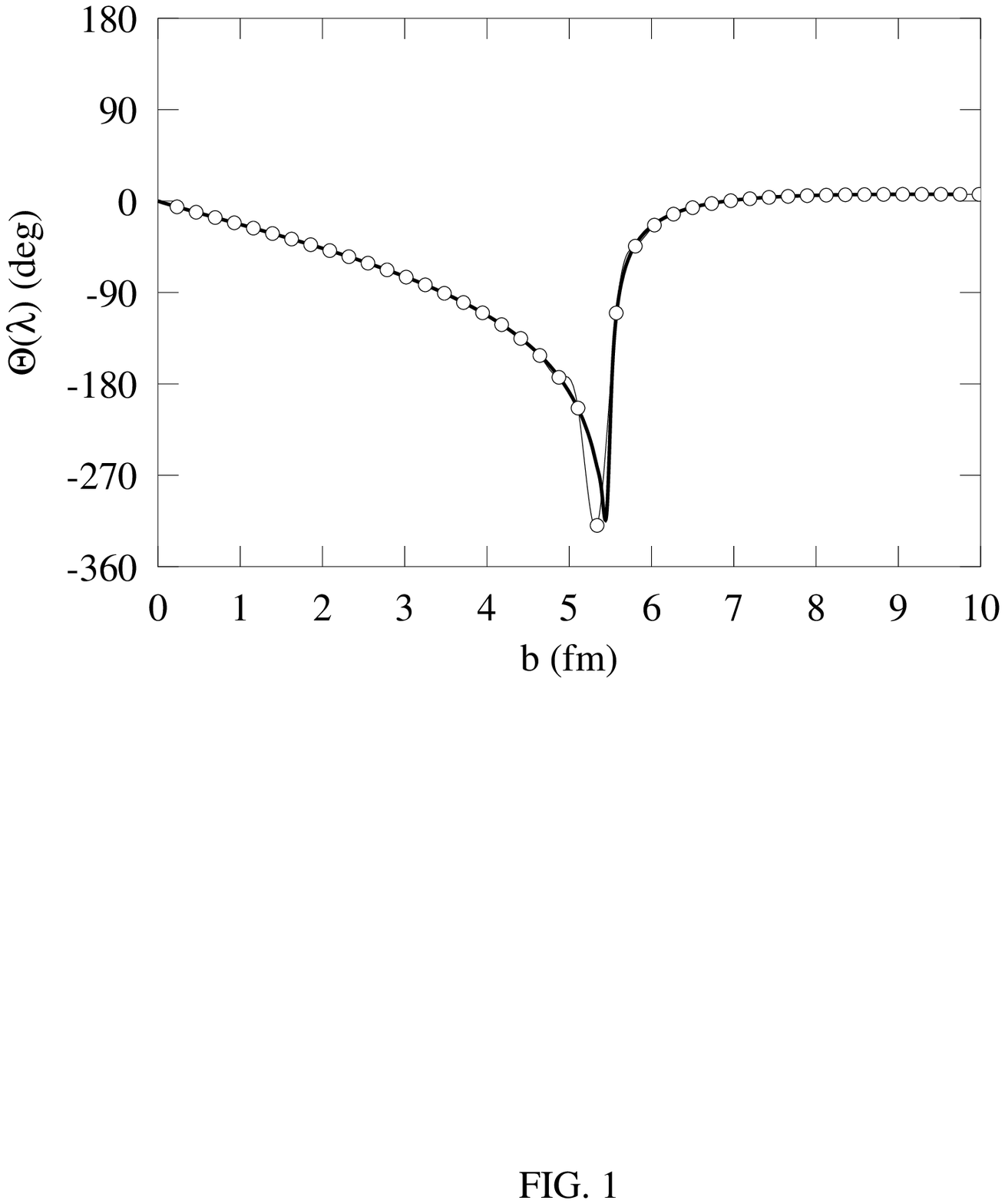,width=8.4cm, clip=}
\caption{The thick curve shows the classical deflection function. 
The open dots represent the values of the quantum deflection function
calculated at integer $\lambda=bk$ values.
The thin curve shows the cubic spline interpolation of the open dots.}
\end{figure}

The agreement between the dots and the thick curve is impressive and
the small differences between the thin and the thick lines may be a
consequence of the method used to interpolate the dots.

The classical deflection function shows a maximum of about $7^\circ$ at
$b_C \simeq 9.3$ ($\lambda_C \simeq 40.2$) and a minimum of about
$-310^{\circ}$ at $b_n \simeq 5.4$ ($\lambda_n \simeq 23.5$).
In the classical cross section we therefore expect two rainbow
singularities at the scattering angles $\theta_C \simeq 7^\circ$
(Coulomb rainbow) and $\theta_n \simeq 50^\circ$ (nuclear rainbow).

Six different branches of $\Theta(\lambda)$ contribute to the cross
section, four corresponding to trajectories  coming from the scattering
half plane containing the scattering angle (near-side trajectories) and
two coming from the opposite half plane (far-side trajectories).

At $b \ne 0$, the deflection function $\Theta(\lambda)$ crosses three
times values for which $\sin \theta =0$. 
Two glory singularities are expected at $\theta=180^\circ$  and one,
additional to the Coulomb singularity, at $\theta=0^\circ$.

In the panel (a) of Fig.2 the thick line shows the ratio of the
classical mechanics cross section to the Rutherford one.
The thin solid and dashed lines give the contributions to the same
quantity from the different branches, near- and far-side respectively,
of the deflection function.

The appearance of the rainbow singularities are manifest, as are the
glory singularities corresponding to the crossing of the deflection
function of the $-180^\circ$ deflection angle.
The glory corresponding to the crossing of the $0^\circ$ deflection
angle is masked by the Rutherford cross section higher order
singularity at the same scattering angle.

In the semiclassical limit a phase factor is associated to each
contribution from the different branches of the deflection function.
In the panel (b) of Fig. 2 the thin solid lines show the interference
limits of these contributions, while the thick dashed and solid lines
give, respectively, the classical and the quantum cross sections.
\begin{figure} 
\label{FIG2}
\hspace*{-3mm}
\epsfig{file=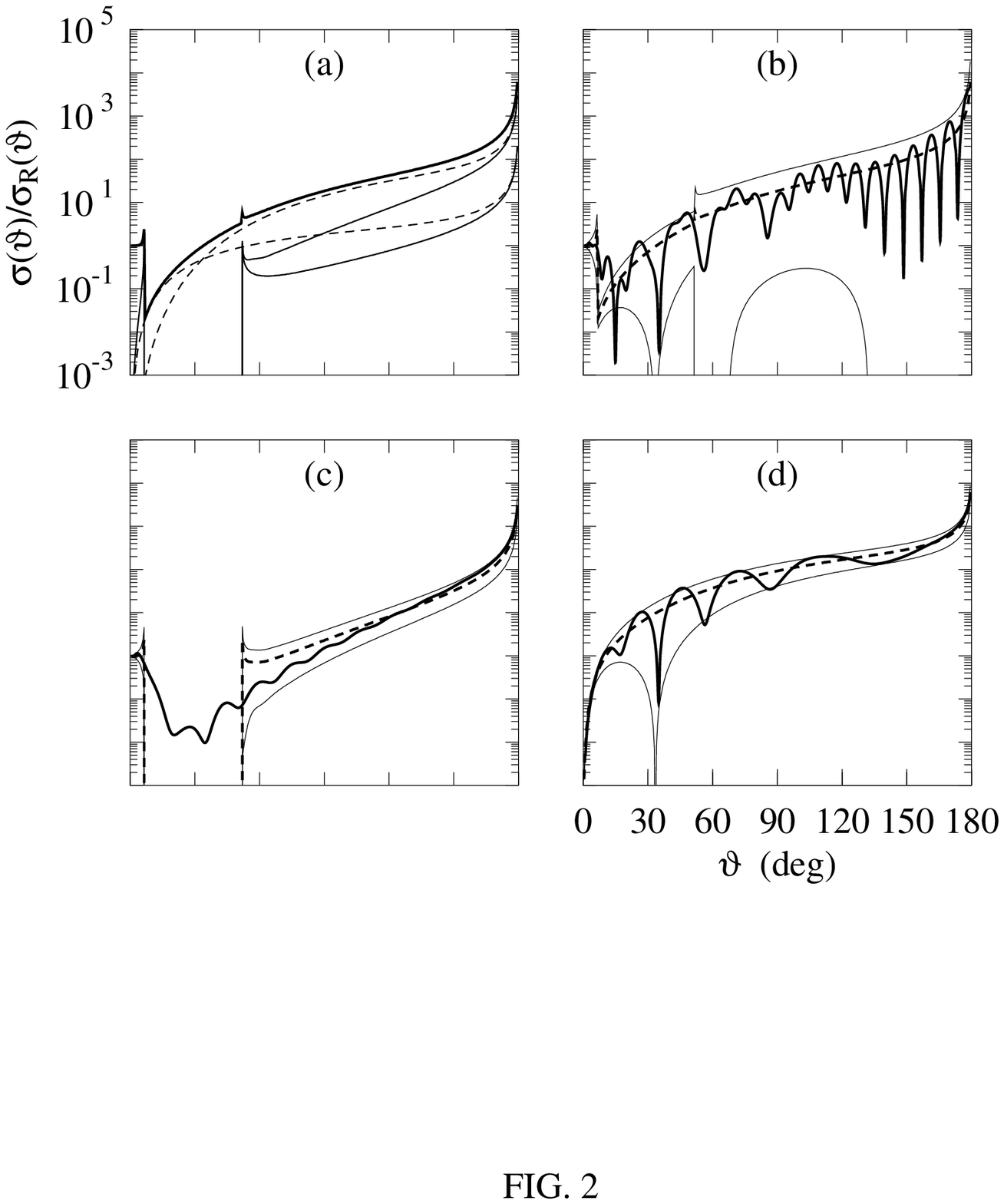, width=8.4cm, clip=}

\caption{ (a) Classical cross section (thick line),  near- (thin solid
lines) and far-side (thin dashed lines) contributions to the classical
cross section; (b) quantum (thick solid line) and classical (thick
dashed line) cross sections , the thin lines show the interference
limits; (c) the same as in panel (b) for the near-side cross sections;
(d) the same as in panel (b) for the far-side cross sections. }
\end{figure}

With the exclusion of a small angular range at the right of $\theta_C$,
the oscillations appearing in the quantum cross section are well within
the interference region.
The interference limits cannot, however, be considered the upper and
lower envelopes of the quantum cross section.
The reason for this is clarified in the panels (c) and (d) of the same
figure, where the thick solid and dashed lines show the near- and
far-side components of the quantum\cite{FUL75} and classical cross
sections, respectively.
The interference limits of the classical far-side cross section are
almost perfect envelopes of the quantum far-side cross section.
The period of the oscillations of this cross section decreases with
the increase of the scattering angle, which corresponds to a decrease of
the deflection angle.
This tendency is confirmed by the very long period of the oscillation
appearing in the backward angles near-side cross section.
Owing to this, the oscillations can be interpreted as arising from
interference of classical-like trajectories whose phase differences
tend to decrease while approaching the nuclear rainbow angular
momentum.
In the present case, these phase differences are too small to allow us
to observe the maximal constructive and destructive interference
amongst all the four branches of the deflection function contributing
to angles larger than $\theta_n$.

In the classical near-side cross section a dark region is present
between $\theta_C$ and $\theta_n$.
From both the shadow boundaries, this dark region appears partially
enlightened by the quantum near-side cross section, and the tails of
the two shadows overlap with each other producing some interference.
The shadow contribution to the right of $\theta_C$ decreases rather
rapidly.
This makes difficult to justify, as arising from this contribution, the
persistence of the interference pattern in the near-side cross section
at angles larger than about $30^\circ$.
In this angular range, the oscillations suggest the existence of an
additional non-classical contribution.

The near- and far-side cross section interference patterns are 
considerably simpler than the full cross section one.
This complicate interference pattern of the full cross section arises
from the coherent superposition of the simpler far- and near-side
amplitudes.
It is the folding of the plane of Fig. 1, necessary to obtain the
dependence of the scattering angle from the impact parameter, which is
responsible for this complicate interference pattern.
The near- and far-side decomposition allow one to unfold the quantum
cross section, considering its dependence on the deflection angle
rather then on the scattering angle.
In Fig. 3 the thick solid line shows this unfolded cross section.
In order to eliminate the appearance of the glory singularities in the
classical cross sections, in this figure the cross sections multiplied
by $\sin \theta$ are plotted.
\begin{figure} 
\label{FIG3}
\hspace*{-3mm}
\epsfig{file=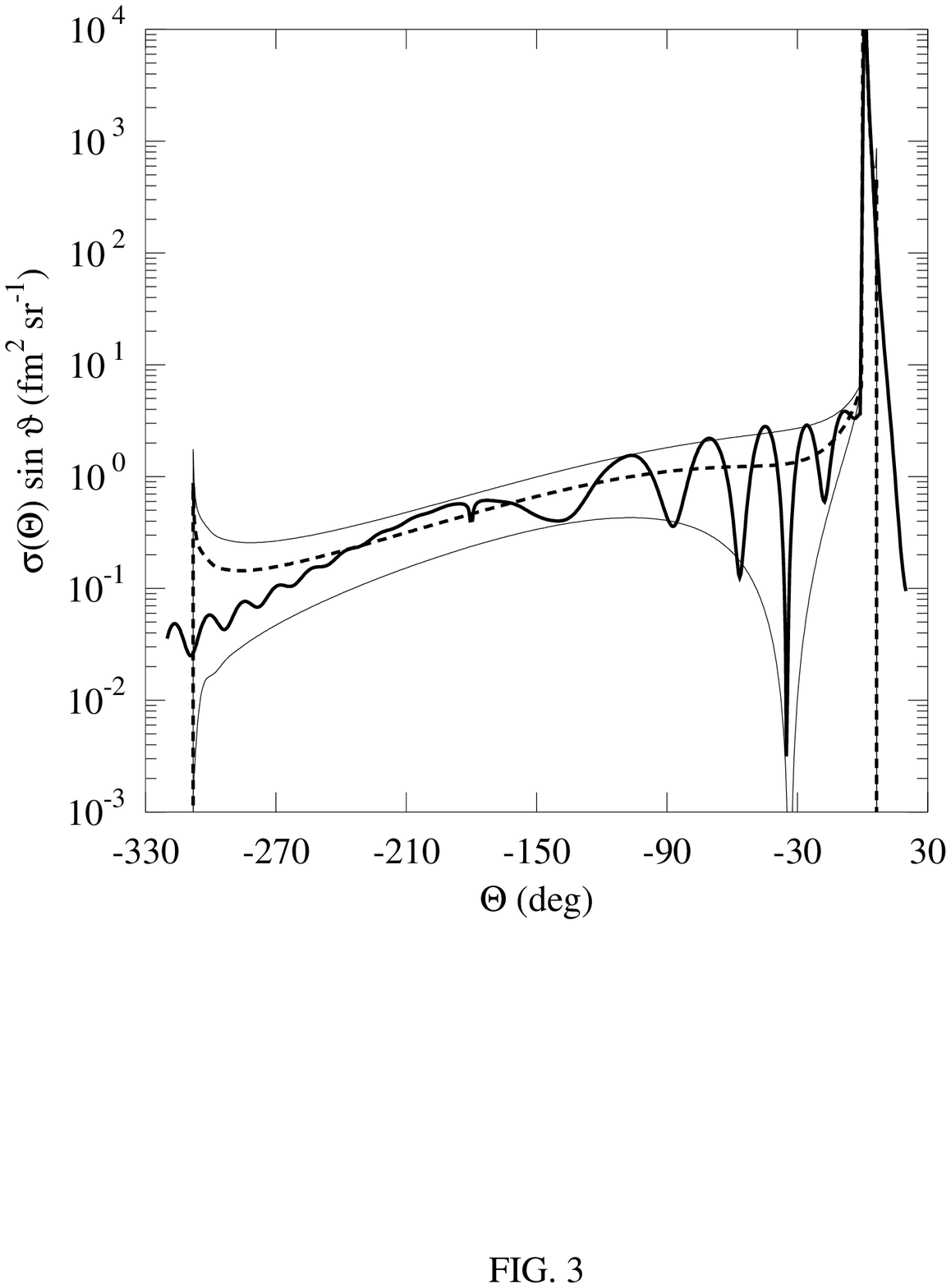, width=8.4cm, clip=}
\caption{Unfolded quantum (thick solid line) and classical (thick
dashed line) cross sections, and classical interference limits (thin
lines), plotted versus the deflection angle.}
\end{figure}

The unfolded quantum curve shows an irregular behavior in a small
angular interval around  $-180^\circ$.
This irregular behavior is probably due to the fact that the
singularities of the quantum near- and far-side cross sections at
$180^\circ$ are slightly different from the $1/\sin \theta$ singularity
predicted by the non-uniform approximation of the Legendre functions.

With the exclusion of this small interval, one can however appreciate
the attempt of the near-side curve ($\Theta < -180^\circ$) to match
continuously the far-side one ($-180^\circ < \Theta < 0^\circ$).

The comparison of $\Theta_Q(\lambda)$ with $\Theta(\lambda)$, and of
the quantum cross sections (full, near- and far-side) with the
corresponding classical ones, allows one to recognize the presence, in
the quantum quantities, of contributions which are very close to those
expected from classical-like trajectories.

A nice Airy-like pattern appears in the unfolded quantum cross section.
The increase of the period of the main oscillations, with decreasing the
deflection angle, well justifies the fact that the interference limits
are rather far from representing the envelopes of the full quantum
cross section.

\subsection{Pure quantum mechanical analysis}

The comparison of the classical and quantum cross sections allows one
to derive clear evidences of classical-like characteristics in the
quantum cross section.
The same result can be obtained without using any classical mechanics
calculation, by simply observing the changes produced in the scattering
function and in the cross section by changing the value attributed to
$\hbar$.

\begin{figure} 
\label{FIG4}
\hspace*{-3mm}
\epsfig{file=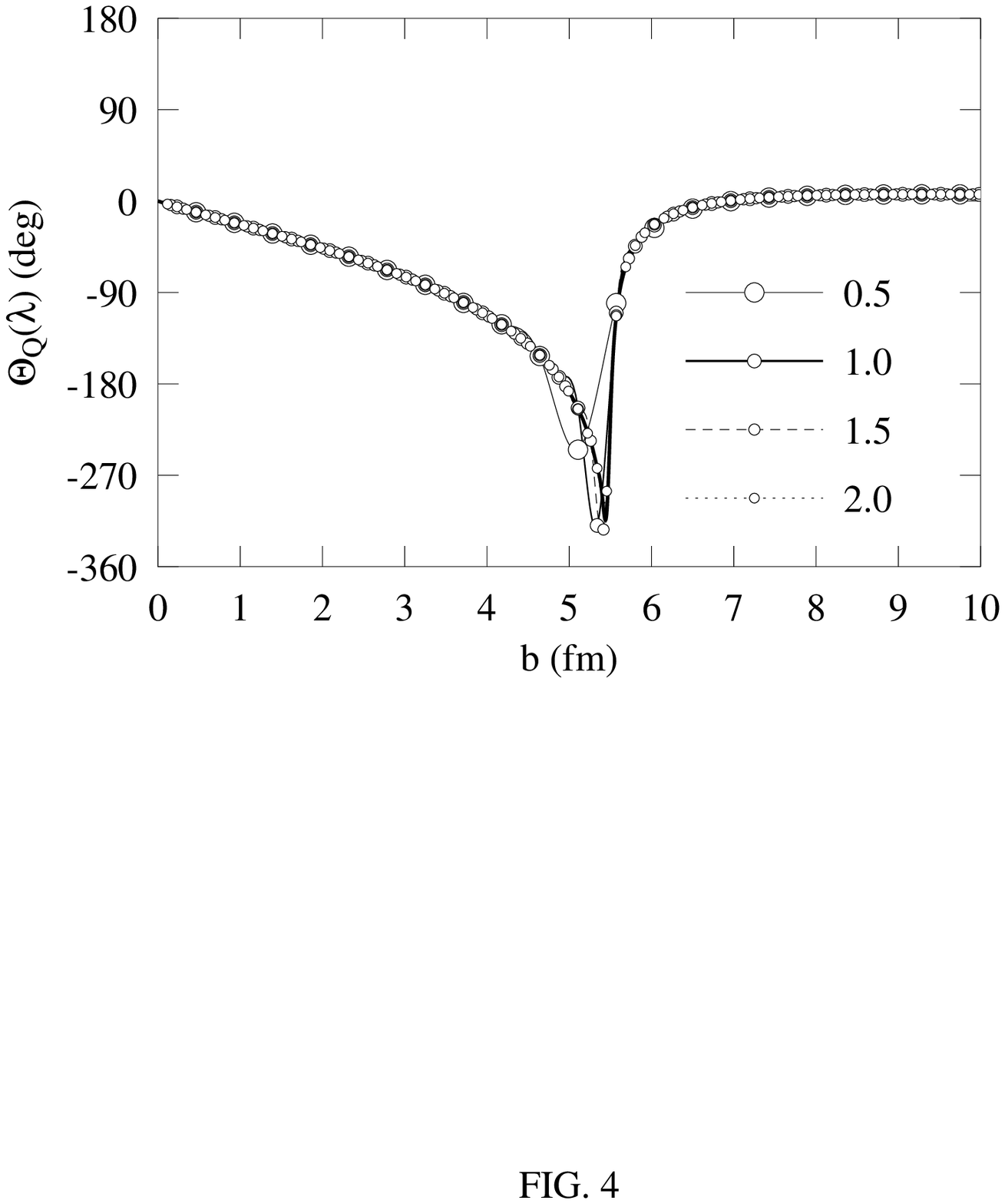, width=8.4cm, clip=}
\caption{ The open dots represent the values of the quantum deflection
functions calculated, at integer $\lambda=bk$ values, for the 4 values
of the $\hbar$ reduction factor given in the figure.
The thin curves show the cubic spline interpolations of the dots and
the thick curve show the classical deflection function.}
\end{figure}

In Fig. 4 the open dots show the values of $\Theta_Q(\lambda)$
calculated at integer $\lambda$ values using Eq. \ref{QuaDef} and
substituting $\hbar$ in an optical potential code with
$\hbar_f=\hbar/f$, with $f=0.5, 1.0, 1.5, 2.0$.

Because the spacing in $b$ of points corresponding to an increment of
one unit in $l$ is proportional to $1/f$, the abscissae of the points
corresponding to $\Delta l = 1, 2, 4$ for the cases $f=0.5, 1.0, 2.0$
are trivially the same at appropriate $b$ values (the b value
corresponding to $\lambda=1$ for $f=0.5$ is the same as that
corresponding to $\lambda=2$ for $f=1.0$ and to $\lambda=4$ for
$f=2.0$, and so on).
In Fig. 4 the open dots corresponding to these values of $f$ result
perfectly concentric at the common $b$ values, with the exclusion of a
small range around $b \simeq 5.5$.
This provides a striking confirmation of the classical scale invariance
properties of $\Theta_Q(\lambda)$ for almost all the values of the
angular momentum.

The thick solid curve, representing $\Theta(\lambda)$, shows that also
in the above small range, and in the more unfavorable case ($f=0.5$,
$\hbar$ two times larger), the agreement between $\Theta_Q(\lambda)$
and $\Theta(\lambda)$ is rather good.
This agreement becomes practically perfect for the most favorable case
($f=2.0$, $\hbar$ two times smaller).

The tendency of all the $\Theta_Q(\lambda)$ points (plotted against $b$
for different $\hbar$ values) to lie on the same curve is a clear
signature of the dominance of the classical dynamics in the
determination of the properties of $S(\lambda)$.
This allows one to obtain information on the classical properties of
$S(\lambda)$ using only quantities calculated by a conventional optical
potential code, and the direct calculation of the classical deflection
function is not really necessary.

In Figs. 5, 6 and 7 are shown the full, near- and far-side cross
sections, respectively, calculated with the values of the reducing
factor $f$ given in the figures.

\begin{figure} 
\label{FIG5}
\hspace*{-3mm}
\epsfig{file=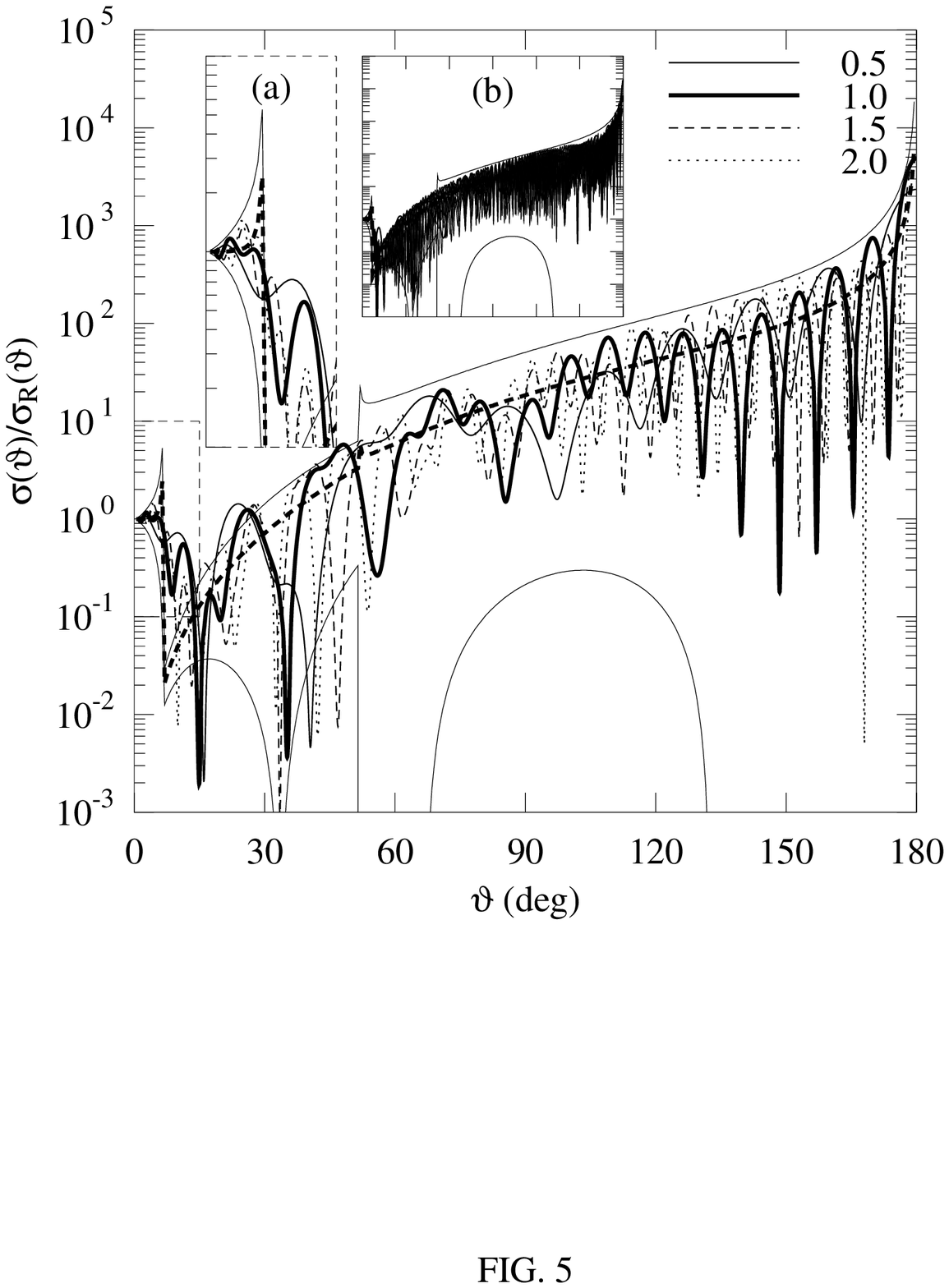, width=8.4cm, clip=}
\caption{Quantum cross sections for the 4 value of the $\hbar$ reduction
factor given in the figure.
The thick dashed and thin solid curves show, respectively, the
classical cross section and the interference limits.
The inset (a) gives an enlargement of the rectangular area of the
figure limited by the dashed lines.
The inset (b) gives a reduction of the whole area of the figure.
In the inset (b) are plotted the quantum, the classical cross sections
and the interference limits together with the quantum cross sections
calculated with values of the reduction factor from 3 to 4.}
\end{figure}

\begin{figure} 
\label{FIG6}
\hspace*{-3mm}
\epsfig{file=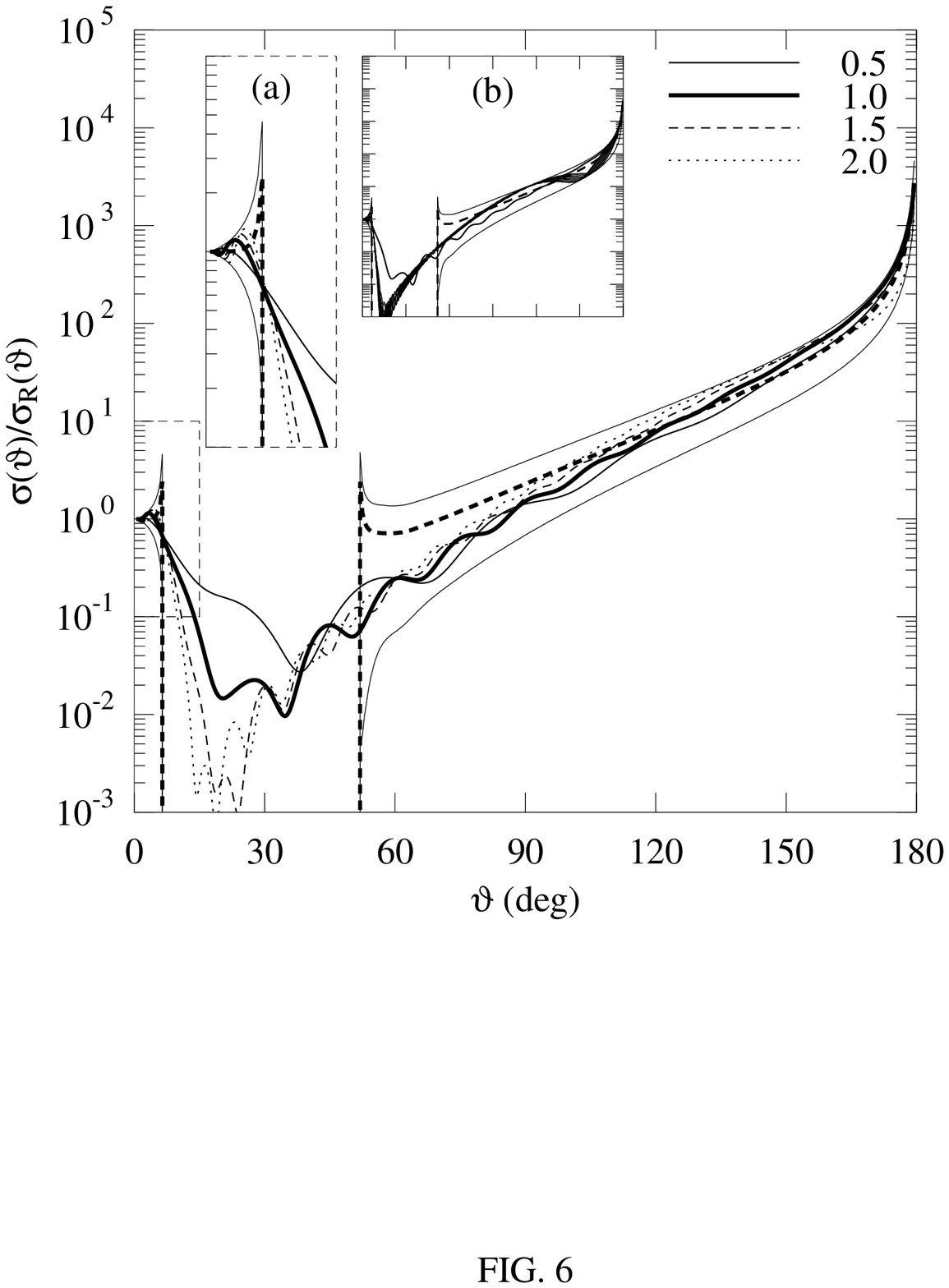, width=8.4cm, clip=}
\caption{The same as Fig. 5 for the near-side cross sections}
\end{figure}

\begin{figure} 
\label{FIG7}
\hspace*{-3mm}
\epsfig{file=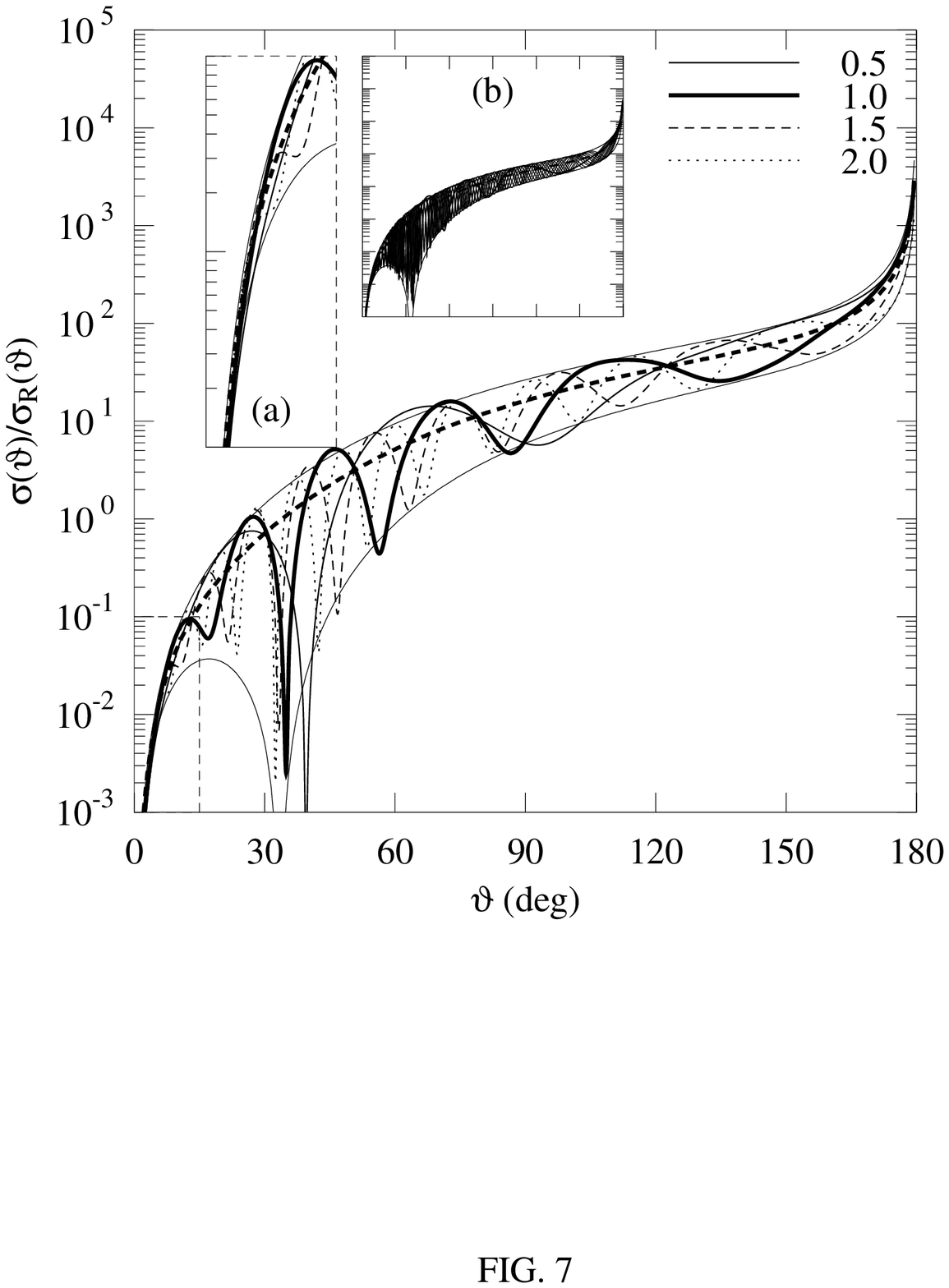, width=8.4cm, clip=}
\caption{The same as Fig. 5 for the far-side cross sections}
\end{figure}

In these figures the thick dashed and thin solid lines show the
classical cross sections and their interference limits, respectively.
These lines were drawn only to remember these classical limits.
As for the deflection function, the drawing of these classical
quantities is not necessary to recognize the presence of classical-like
contributions.

In all these figures the insets (a) show an enlargement, by a factor 2,
of the rectangular regions delimited by the dashed lines, and the
insets (b) show a reduction, by a factor 3, of the complete figures.
In the insets (b) are plotted the true cross sections together with
eleven cross sections calculated with values of $f$ ranging from 3.0 to
4.0 with a step of 0.1.

By looking at Fig. 5 one observes a rather complicate behavior of the
cross sections corresponding to the four values of $f$ considered.
This makes it difficult to imagine that the oscillations tend to be
confined within well defined regions. 
This tendency begins to appear in the inset (b), where a rather well
defined upper envelope can be observed, and also indications of a lower
envelope are present.
One explanation of the minor definiteness of the lower envelope can be
found in the fact that, with the scale used, the minima are much
narrows than the maxima.
Using a fixed grid to tabulate the cross section it is more probable to
miss a minimum rather than a maximum.

In the angular region delimited by $\theta_C$ and $\theta_n$, the full
cross sections calculated with the four values of $f$ is in
disagreement with the interference limits, particularly in the region
to the right of $\theta_C$.
This disagreement decreases rapidly with increasing $f$.

The reason of this behavior is understood by considering Fig. 6, were
the near-side cross sections are plotted.
In the classical shadow region, by increasing $f$, the cross sections
decrease very rapidly moving to the right of $\theta_C$, while they
decrease slowly moving to the left of $\theta_n$.
In the inset (b) one can observe that even for $f$ values ranging from
3.0 to 4.0 the decrease of the near-side cross section is slow, moving
from the right towards the shadow region.
The eleven cross sections only begin to fill gradually the region
defined by the interference limits. The phase difference between the
classical-like trajectories, which contribute to this part of the
near-side cross section, depend weakly on the angle and only a few
oscillations appear in the cross section at the maximum value of $f$
considered.

In the treatment of the scattering amplitude using the uniform method
around a rainbow angle, the rapidity of the decrease of the cross
section, in the classical shadow region, depends on the second
derivative of the deflection function at the rainbow angular momentum.
The curvature of the deflection function is much higher at the nuclear
than at the Coulomb rainbow, and this explains why the two slopes are
so different.

Comparing the inset (a) of Fig. 6 with the corresponding inset of Fig.
5, one can also observe that the interference pattern of the near-side
cross section, around $\theta_C$, is considerably simpler than that of
the full cross section.
The additional oscillations in the full cross section arise from the
contributions from the far-side amplitude.

The full and the near-side cross sections do not clearly exhibit
properties which are invariant with respect to the value attributed to
$\hbar$.
This is not the case for the far-side cross sections which are given in
Fig. 7.
The existence of common upper and lower envelopes for these cross
sections is rather well indicated already by the $f$ values ranging
from 0.5 to 2.0, and is clearly proved by the $f$ values from 3.0 to
4.0 given in the inset (b) of the figure.
Apart from a small distortion of at least one of the two interfering
amplitudes this figure provides a strong indication of the dominance of
the contributions from classical-like trajectories already from the
value $f = 0.5$ ($\hbar$ two times larger).

\section{Complex optical potential cross section: $E_{\rm Lab}=132$ MeV }

The first complex optical potential considered is one of the potentials
whose cross section fits the experimental data at $E_{\rm Lab} = 132$
MeV\cite{OGL00}.
The imaginary part of the potential has a conventional Woods-Saxon form
factor with parameters $W_{0}=13.86$ MeV, $R_{w}=5.6894$ fm, and
$d_{w}=0.656$ fm.
Its real part is that of the real optical potential previously
considered.

This case was chosen because a recent semiclassical
analysis\cite{ANN01}, using the Brink and Takigawa\cite{BRI77}
approximation, has shown that the oscillations appearing in the
far-side cross section can be explained as arising from the
interference between far-side contributions from the first two terms of
the multireflection expansion of the semiclassical scattering
amplitude\footnotemark.
Because the far-side contribution to the barrier term is responsible
for the appearance of the Fraunhofer-like pattern in the barrier cross
section, one is naturally induced to think that this contribution
should be considered of diffractive nature, i.e. of quantum origin.

\footnotetext{In the following, the first two terms of the
multireflection expansion of the Brink and Takigawa scattering function
will be named, respectively, {\it barrier} and {\it internal} terms.
This terminology, which is rather common in the literature, will be
adopted for simplicity, in spite of the presence of contributions from
the internal part of the interaction also in the higher order terms of
the multireflection expansion.}

It seems therefore interesting to test if the simple recipe here
proposed is able to discover the non-classical origin of this
contribution.
By decreasing the value of $\hbar$ and approaching the classical
mechanics limit, one should also observe how this diffractive
contribution becomes a classical one.

\subsection{Comparison with the classical cross section}

According to the above classical interpretation of the imaginary part
of the potential, the presence of this term does not modify the
classical deflection function $\Theta(\lambda)$.
The imaginary part only introduces a probability $P(\lambda)$ of
survival in the elastic channel for particles with angular momentum
$\lambda \hbar$.

In the panels (a) and (b) of Fig. 8 the thick lines show, respectively,
the square root of $P(\lambda)$ and $\Theta(\lambda)$ as functions of
the impact parameter $b$.

\begin{figure} 
\label{FIG8}
\hspace*{-3mm}
\epsfig{file=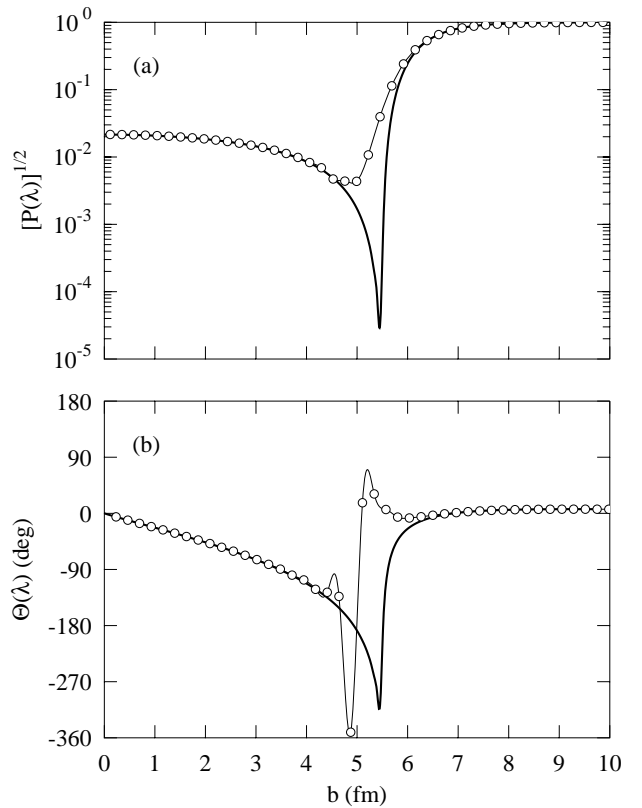, width=8.4cm, clip=}
\caption{The thick curves show the classical deflection function (panel
(b)) and the square root of the survival probability (panel (a)).
The open dots represent the values of the quantum deflection function
and of the modulus of the scattering function, calculated at integer
and half integer $\lambda=bk$ values, respectively.}
\end{figure}

In the same figure the dots represent the values of $|S(\lambda)|$ and
of $\Theta_Q(\lambda)$\footnotemark and the thin lines are cubic spline
interpolations of the calculated values.
The figure shows that the quantum and classical corresponding
quantities are in good agreement, apart from a neighborhood around $b
\simeq 5$ fm of half width of about 1 fm.
The impact parameter value $b_n$ of the nuclear rainbow is in this
region, and its position is very close to the position of the deep
minimum of $P(\lambda)$.
The behavior of $S(\lambda)$ in this region is considerably different
from what would be expected in the extreme classical limit.
This suggests the dominance of a scattering mechanism different from
the classical one for particles with angular momenta corresponding to
this range of impact parameter values.

\footnotetext{The quantum deflection function shown in Fig. 1 of
Ref.\cite{ANN01} was obtained by using for all the $\lambda$ values an
increment $\Delta \lambda=\frac{1}{2}$ in Eq. \ref{QuaDef}.
The comparison of this figure with our Fig. 8 shows that even this
rough method produces a correct estimate of the quantum deflection
function with the exclusion of only one $\lambda$ value.
This value corresponds to a very large variation of $\arg S(\lambda)$
between two consecutive integer $l$ values.
}

\begin{figure} 
\label{FIG9}
\hspace*{-3mm}
\epsfig{file=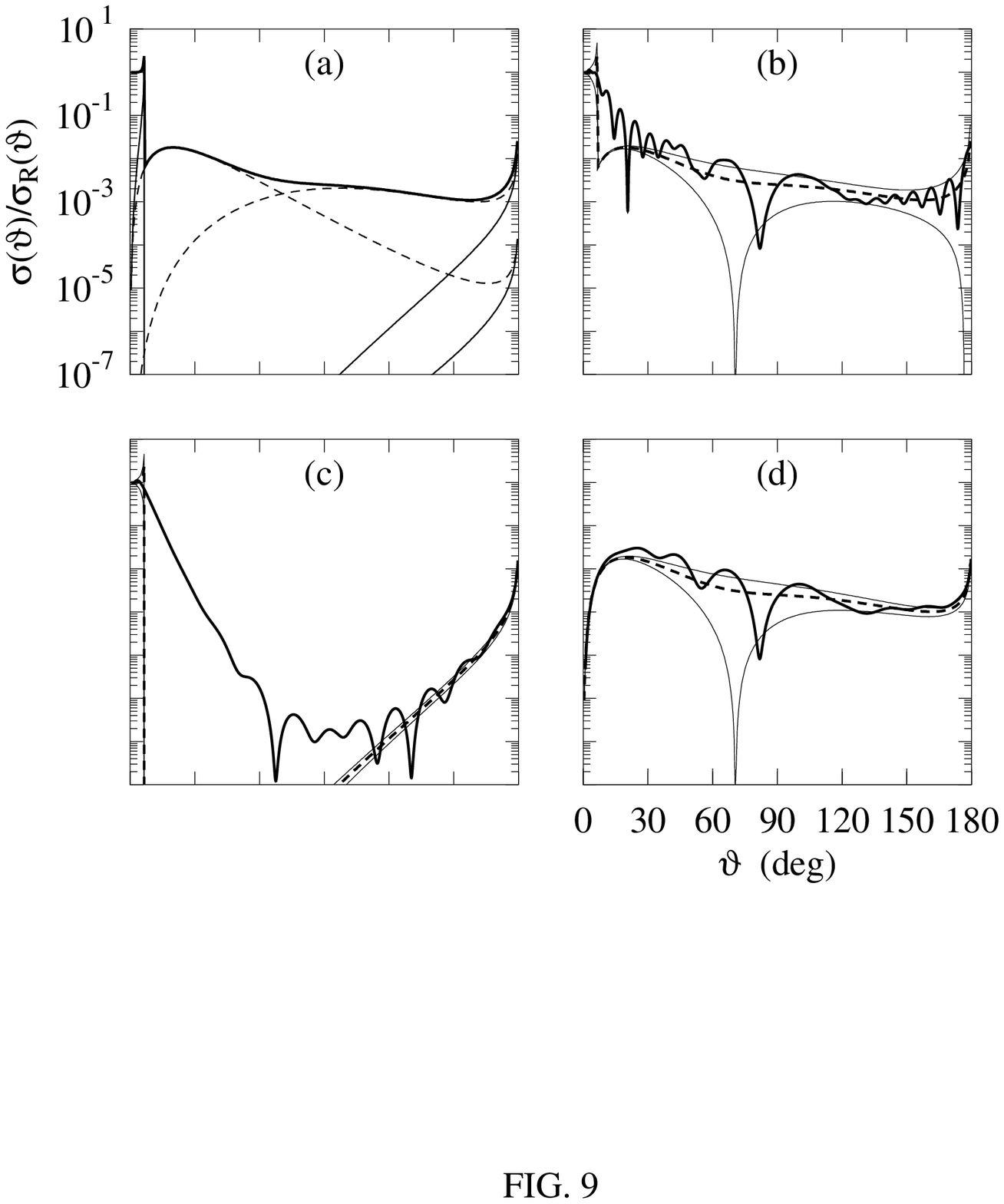, width=8.4cm, clip=}
\caption{The same as Fig. 2 for the complex optical potential at
$E_{\rm Lab}=132$ MeV.}
\end{figure}

In the different panels of Fig. 9 are shown the same quantities given
in the corresponding panels of Fig. 2 for the real potential.
At angles larger than $\theta_C$ the contributions from all the branches
of the deflection function are strongly reduced by the absorption.
Each point of the old curves is lowered by the corresponding value 
of $P(\lambda)$.
Furthermore, the deep minimum of $P(\lambda)$, around $b_n$, produces a
dramatic reduction of the contribution from the near-side trajectories
with $b \simeq b_n$.
Only part of the near-side contributions from the corresponding two
branches of $\Theta(\lambda)$ (the two thin solid lines in the right
bottom  corner of panel (a)) can be observed within the range of the
vertical axis of Fig. 9.
At backward angles these curves are close to the dashed lines
representing the far-side contributions, but they drop very rapidly
with decreasing angle going out from the plotted area.
The very small values of these contributions at angles just above
$\theta_n$ prevent the observation of effects in the classical cross
section deriving from the nuclear rainbow singularity.

The rapid decrease of these contributions, together with the
modifications of the slopes of the contributions from the far-side
trajectories, considerably shrinks the width of the interference region.
The borders of this region are shown by the thin lines in the panel (b)
of Fig. 9.
In the same panel, the thick curve shows the quantum cross section.
This curve, in the forward hemisphere, substantially violates the
boundaries fixed by the interference limits.

In the panel (c) we show that the violation of the interference limits
is mainly due to a violation of the corresponding limits by the
near-side component of the full cross section.
At angles to the right of $\theta_C$ the quantum curve decreases almost
exponentially, at the rate of about one order of magnitude per
$10^\circ$, filling the classical shadow region between $\theta_C$ and
$\theta_n$.
By increasing the angle, oscillations of increasing amplitudes appear,
indicating the interference of the exponentially decreasing
contribution with another one.

In the backward hemisphere, decreasing the angle below $180^\circ$, the
quantum near-side curve initially follows the classical one, then it
begins to show oscillations with an amplitude increasing with
decreasing angle.
These oscillations can be interpreted as arising from the interference
of a classical-like contribution with a different contribution, of
non-classical origin.
This additional contribution appears to be the same producing the
oscillations in the exponential tail at the right of $\theta_C$.

This behavior is qualitatively similar to that observed in the
near-side cross section of the real potential.
The only difference is represented by the fact that now the curves at
the right of $\theta_n$ are downward shifted and decrease more rapidly
by decreasing the angle.
This allows one to observe in a wider angular range the exponential
decrease at the right of $\theta_C$.

The quantum and classical far-side cross sections given in the panel
(d) show that a relevant contribution to the violation of the classical
interference limits comes also from at least one of the two terms
responsible for the oscillatory pattern of the quantum far-side cross
section.
This confirms the results obtained with the semiclassical
analysis\cite{ANN01}, suggesting that one of these contributions is not
of classical origin.

\subsection{Pure quantum analysis}

The panels (a) and (b) of Fig. 10 show, respectively,  $|S(\lambda)|$
and $\Theta_Q(\lambda)$ for the four values of $f$ given in the figure.
By increasing $f$ from 0.5 to 2.0,  all the points tend to lie on the
same curve in increasing ranges of $b$.

Comparing the behavior of the points representing $\Theta_Q(\lambda)$ 
with the corresponding ones for the real potential one observes that
the imaginary part of the potential delays the approach of the
classical limit.
The addition of an imaginary part increases the non-homogeneity of the
medium in which the particles propagate and favors the survival of wave
effects.

\begin{figure} 
\label{FIG10}
\hspace*{-3mm}
\epsfig{file=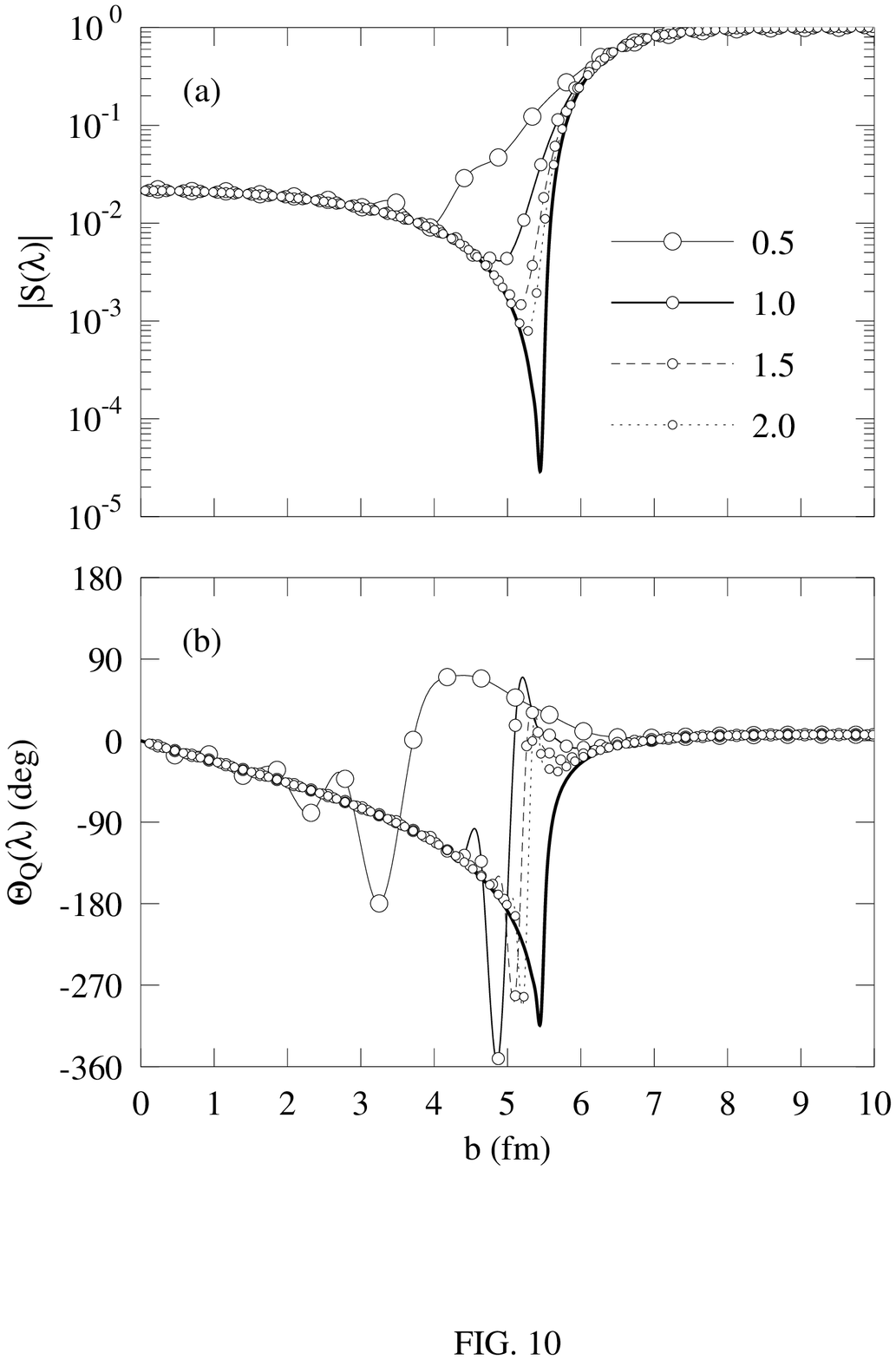, width=8.4cm, clip=}
\caption{ The open dots give the moduli of the scattering functions
(panel (a)) and the quantum deflection functions (panel (b)) calculated,
for the 4 values of the $\hbar$ reduction factor given in the figure,
at half-integer and at $\lambda=bk$ values, respectively.
The thick curves show the corresponding classical quantities and the
other curves are the cubic spline interpolations of the dots.}
\end{figure}

It is interesting to observe that, for $f=0.5$ and $b$ larger then
about 4 fm, $\Theta_Q(\lambda)$ has characteristics typical of a
repulsive interaction.
These are similar to those of the deflection function of the barrier
term of the Brink and Takigawa approximation, which accounts for the
contribution from the reflection phenomenon of the incoming waves in
the region of rapid variation of the properties of the interaction.

The variations with $f$ of the full, near- and far-side cross sections
are shown in Figs. 11, 12, and 13, respectively.

In these figures the curves corresponding to the values of $f$ from 0.5
to 2.0 are rather far from having common envelopes.
In the backward hemisphere and for the full and far-side  cross
sections, these common evelopes only begin to appear for the curves
with $f=1.5$ and $2.0$.
On the contrary, the existence of a well defined interference region is
clearly shown by the calculations with values of $f$  from 3.0 to 4.0,
given in insets (b).

Thanks to the rapid decrease of the near-side cross section, by
decreasing the angle below $180^\circ$, the boundaries of the
interference region are far better defined for the complex potential
full cross section than for the real potential one.
The addition of the imaginary part to the optical potential has
strongly increased the average slope of the backward near-side cross
section and has considerably reduced, or perhaps eliminated, the long
period oscillations appearing in the inset (b) of Fig. 9.
Both these facts contribute to a better definition of the interference
region for the complex optical potential full cross section.

\begin{figure} 
\label{FIG11}
\hspace*{-3mm}
\epsfig{file=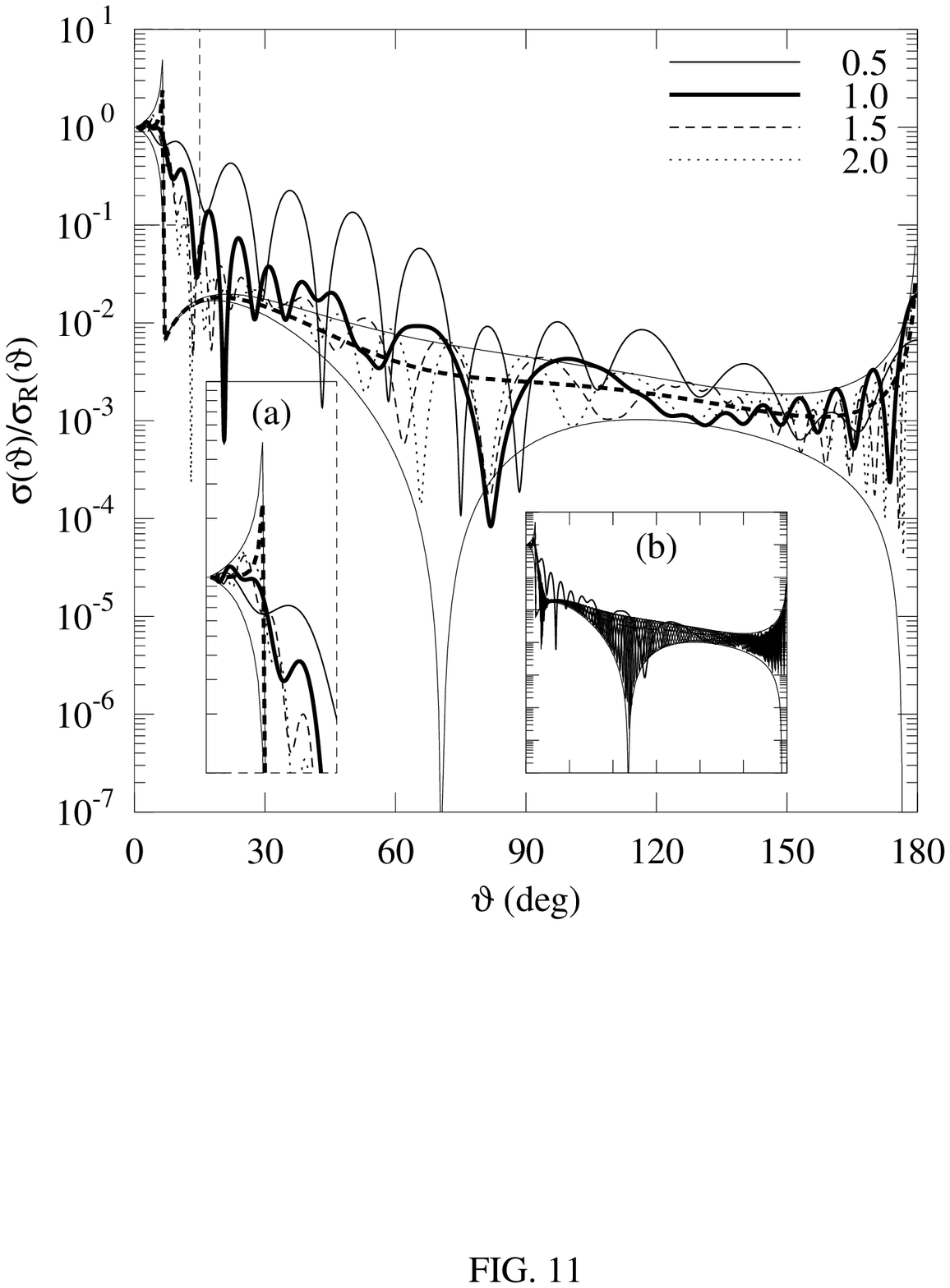, width=8.4cm, clip=}
\caption{The same as Fig. 5 for the full cross sections of the complex
optical potential at $E_{\rm Lab}=132$ MeV.}
\end{figure}

\begin{figure} 
\label{FIG12}
\hspace*{-3mm}
\epsfig{file=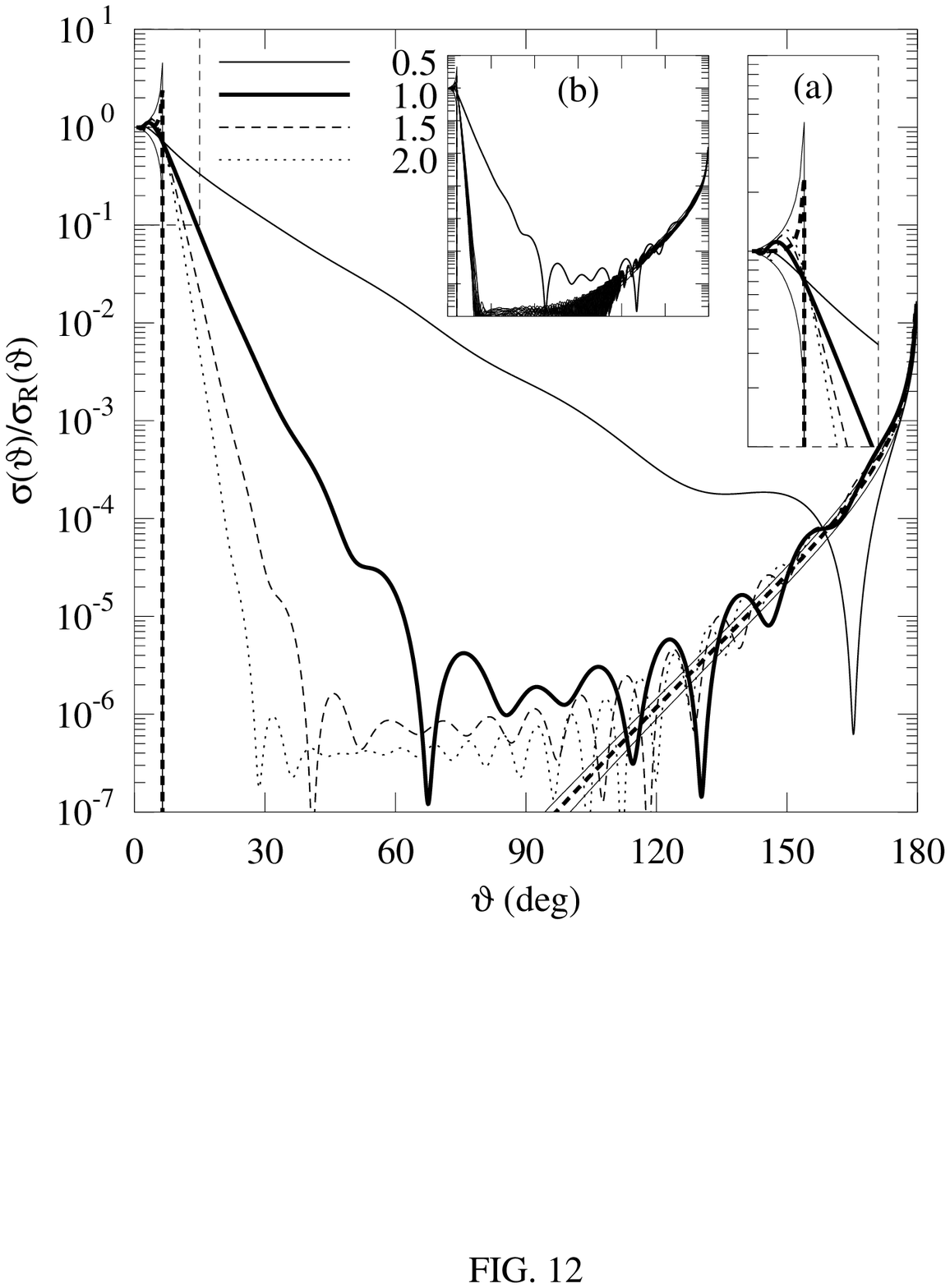, width=8.4cm, clip=}
\caption{The same as Fig. 5 for the near-side cross sections of the
complex optical potential at $E_{\rm Lab}=132$ MeV.}
\end{figure}

\begin{figure} 
\label{FIG13}
\hspace*{-3mm}
\epsfig{file=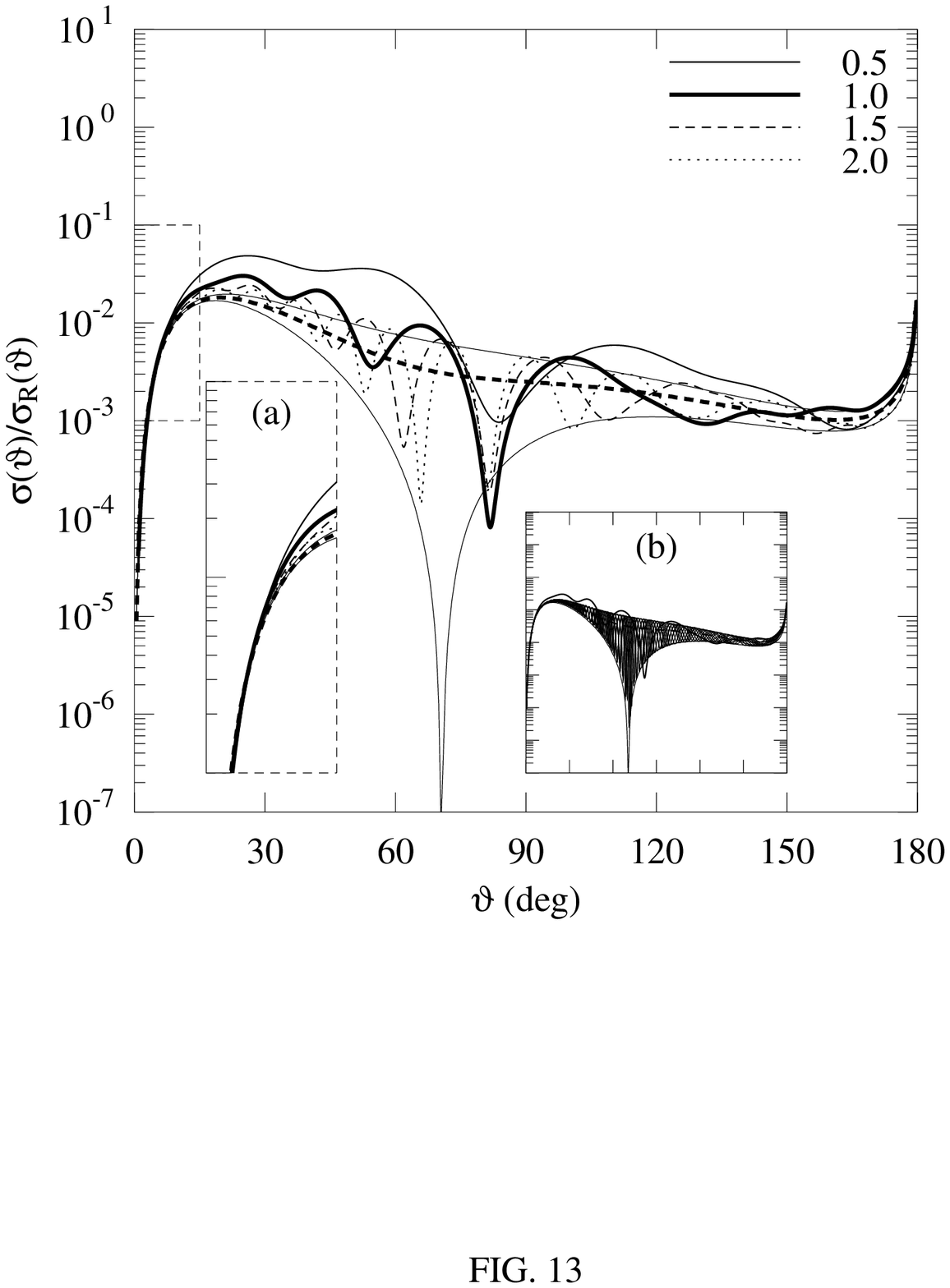, width=8.4cm, clip=}
\caption{The same as Fig. 5 for the near-side cross sections of the
complex optical potential at $E_{\rm Lab}=132$ MeV.}
\end{figure}

The interference region, obtained using a pure quantum calculation, 
is just the one previously calculated using the classical mechanics.
This shows that the analysis of the nature of the different
contributions to the cross section can be done in absence of any
classical mechanics calculation.
 
The presence of at least one classical-like contribution in the far-side
cross section is proved by the behavior of the cross section at
backward angles, and by its continuation, passing through the glory
singularity, in the near-side cross section at backward angle.
At forward angles, the violation of the interference limits  confirms
the non-classical origin of the other contribution, responsible for the
oscillatory pattern in the far-side cross section.

In the classical near-side shadow region, the existence of an
approximately constant contribution to the quantum cross section is
clearly displayed by the dashed and dotted curves of Fig. 12.
This is the contribution which is responsible for the appearance of
oscillations around the exponential drop of the near-side cross section
at the right of $\theta_C$,  and around the classical-like contribution
in the backward hemisphere.
We did not attempted to investigate whether this contribution arises
from some physical fact of from the numerical procedure used to
calculate the cross sections.

\section{Complex optical potential cross section: $E_{\rm Lab}=200$ MeV}

The second complex optical potential considered is one of the
potentials whose cross section fits the experimental data at $E_{\rm
Lab} = 200$ MeV\cite{OGL00}.
This potential also has conventional Woods-Saxon form factors with
parameters $V_{0}=216.3$ MeV, $R_{v}=3.2847$ fm, $d_{v}=0.927$ fm, for
the real part, and $W_{0}=17.83$ MeV, $R_{w}=5.8625$ fm, and
$d_{w}=0.541$ fm for the imaginary part.

This case was considered because an analysis similar to that of
Ref.\cite{ANN01} shows that the semiclassical Brink and Takigawa method
fails to reproduce quantitatively the optical cross section in the
whole angular range.

The results of the semiclassical analysis are summarized in Fig. 14,
where the medium and heavy solid lines show the full semiclassical  and
the exact cross sections, respectively, while the thin solid and dashed
lines show the barrier and the internal cross sections. In the inset (a) 
the corresponding far-side cross sections are shown.

\begin{figure} 
\label{FIG14}
\hspace*{-3mm}
\epsfig{file=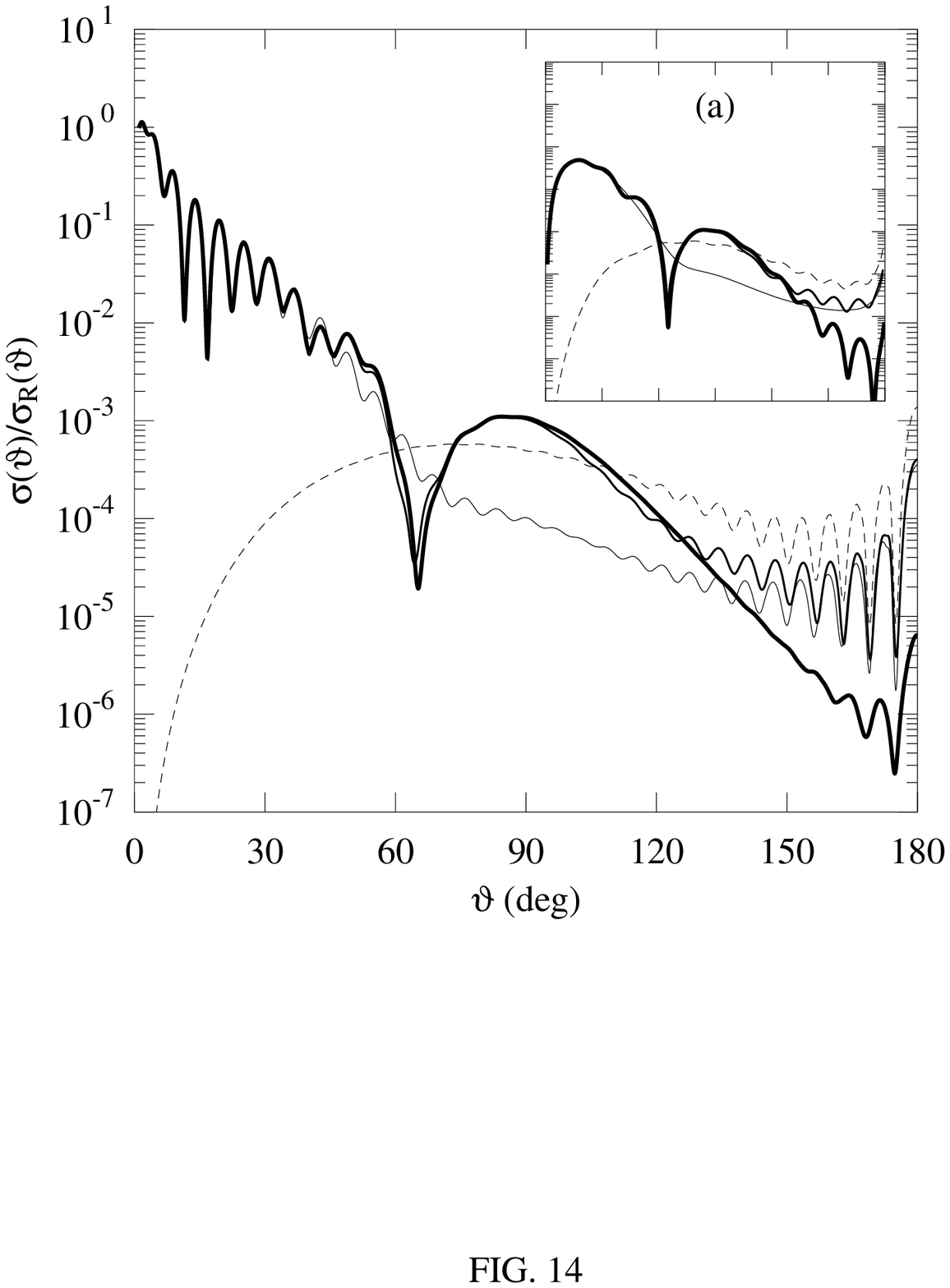, width=8.4cm, clip=}
\caption{ Full semiclassical (medium thick line) and quantum cross
sections (heavy thick line) for the $E_{\rm Lab}=200$ MeV case,
together with the barrier (thin solid line) and the internal (thin
dashed line) cross sections. Using the same lines, in the inset (a)
the corresponding far-side cross sections are shown. }
\end{figure}

The agreement between the semiclassical and the exact cross sections in
rather good only for scattering angles smaller then about
$120^{\circ}$.
As for the $132$ MeV case, in this region the oscillations  appearing
in the exact far-side cross section can be explained as arising from
the interference between far-side contributions from the barrier and
the internal amplitudes.
This casts doubts on the appropriateness of the use of the Airy
terminology for the interference pattern for the $200$ MeV potential.

At angles larger than $120^\circ$, however,  the semiclassical 
and the exact cross sections are in large disagreement and this
disagreement suggests caution in attributing a physical meaning
to the semiclassical analysis.

The reason of the failure of the Brink and Takigawa approximation is
probably connected with the fact that the case here considered is outside
the range of applicability of the method.
For this potential the complex orbiting angular momentum at which the
barrier turning points coalesce is far from the real $\lambda$ axis.
On the contrary a different orbiting angular momentum is very close to
the physical region.
This is the angular momentum at which the internal turning point
coalesces with a turning point different from those usually considered
in the Brink and Takigawa approximation.
This orbiting point, not treated correctly by the approximation, may be
responsible for the anomalous behavior of the semiclassical cross
sections.

The hope is that the present recipe can provide some useful and more
clear indications on the nature of the amplitudes contributing to the
cross section of this potential.

\subsection{Comparison with the classical cross section}

In Fig. 15 the classical $\Theta(\lambda)$ and $\sqrt{P(\lambda)}$ are
shown together with the corresponding quantum quantities. 
With respect to the 132 Mev case, the minimum of the deflection
function corresponding to the nuclear rainbow has moved to a deflection
angle $\Theta_n \simeq -125^\circ$.

Because the nuclear rainbow singularity slids\cite{MCV86} toward a
deflection angle larger then $-180^\circ$, the backward glory
singularities are suppressed, and only four branches of the deflection
function contribute to the cross section.
The panels of Fig. 16 show that, in this case, the regions to the right
of the Coulomb rainbow $\theta_C$ (for the near-side cross section) and
to the right of  the nuclear rainbow $\theta_n$ (for the far-side and
the full cross sections) are classical shadow regions.

\begin{figure} 
\label{FIG15}
\hspace*{-3mm}
\epsfig{file=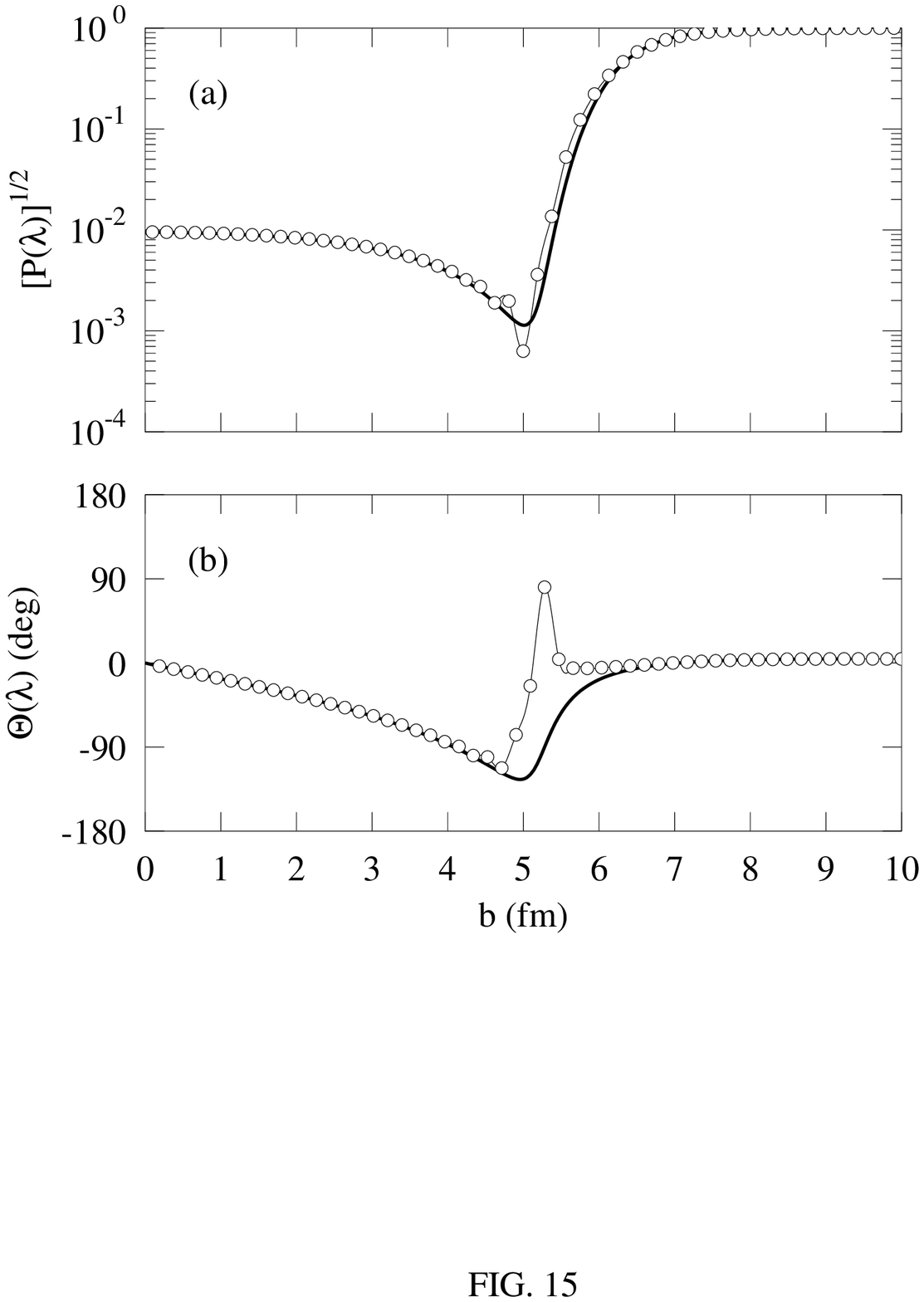, width=8.4cm, clip=}
\caption{The same as Fig. 8 for the complex optical potential at
$E_{\rm Lab}=200$ MeV.}
\end{figure}

The quantities $\Theta_Q(\lambda)$ and $|S(\lambda)|$ are in
substantial agreement with the corresponding classical ones in $b$
ranges wider than for the 132 Mev case.
However, the violations of the interference limits of the quantum full
and far-side cross sections, shown in the panels (b) and (d) of Fig.16,
are not smaller than in the lower energy case.
For angles smaller than about $60^\circ$ the far-side quantum cross
section is largely outside of the classical interference region.
This suggests that also for this potential, as for the 132 MeV case,
these oscillations cannot be interpreted as arising from the
interference between two classical-like contributions.

\begin{figure} 
\label{FIG16}
\hspace*{-3mm}
\epsfig{file=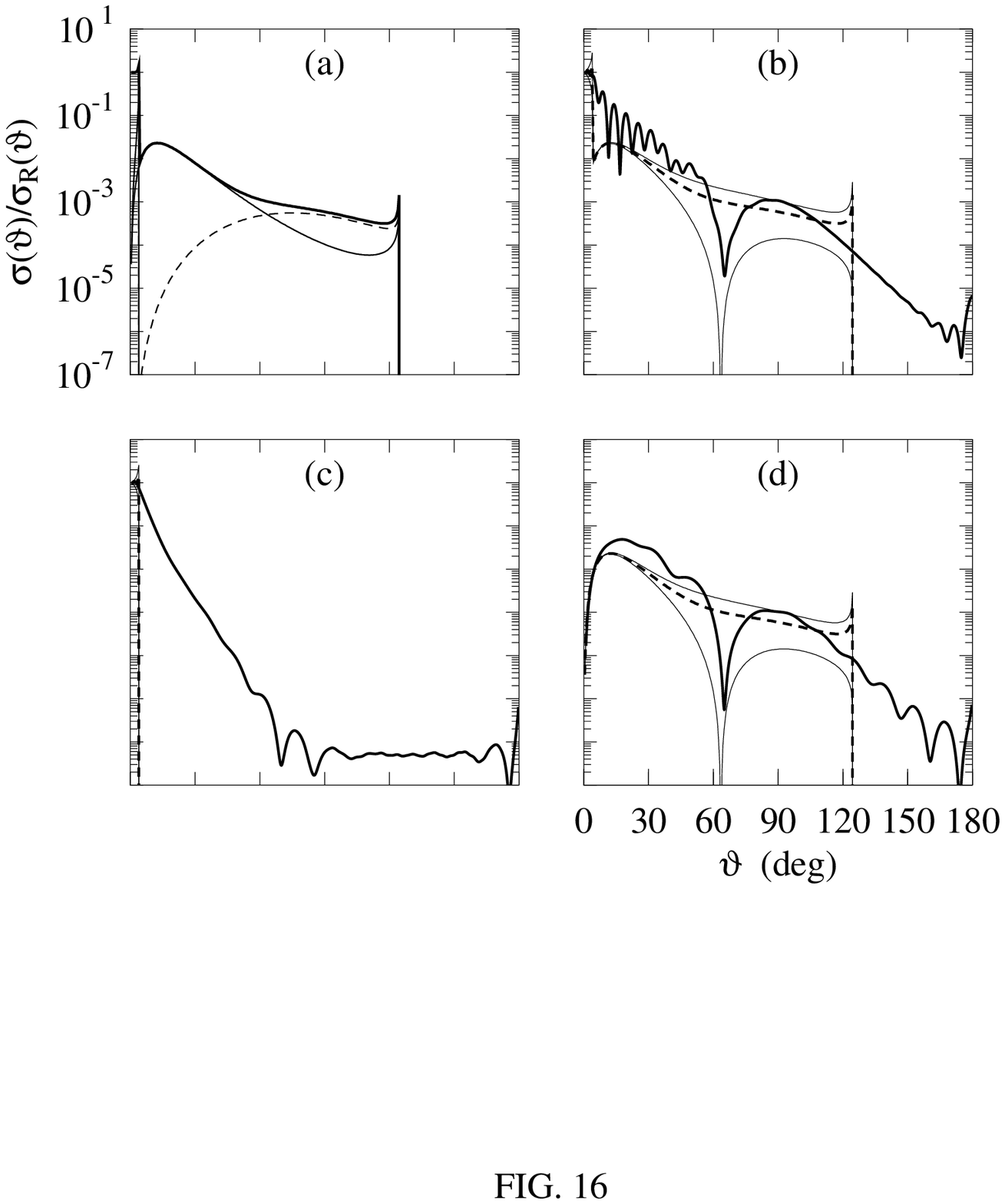, width=8.4cm, clip=}
\caption{The same as Fig. 2 for the complex optical potential at
$E_{\rm Lab}=200$ MeV.}
\end{figure}

In panel (c) of Fig. 16 we clearly see the contribution from an almost
constant additional term, which interferes with the exponential-like
decrease of the cross section to the right of $\theta_C$.
It is more evident than in the corresponding panel of Fig. 9,  where
its presence was barely apparent, in the backward hemisphere, through
its interference with the near-side classical-like contribution in this
region.

\subsection{Pure quantum analysis}

For the 200 MeV potential, the figures from 17 to 20 correspond to the
figures from 10 to 13, for the 132 MeV case.
By comparing Fig. 17 with Fig. 10 one observes that, by decreasing
$\hbar$, the properties of $S(\lambda)$ approach of the classical 
limit faster in the higher energy case.
This is true also for the properties of the cross sections, and depends
on the fact that for the higher energy the wavelength is smaller.
In particular, by looking at Fig. 18 and Fig. 20 one observes that the
quantum curves begin to have as upper and lower envelopes the
interference limits, for scattering angles around $60^\circ$, already
with a $\hbar$ reducing factor of 1.5.
From this value upward the interference pattern, below the classical
nuclear rainbow angle, can be considered a genuine Airy-like pattern.

\begin{figure} 
\label{FIG17}
\hspace*{-3mm}
\epsfig{file=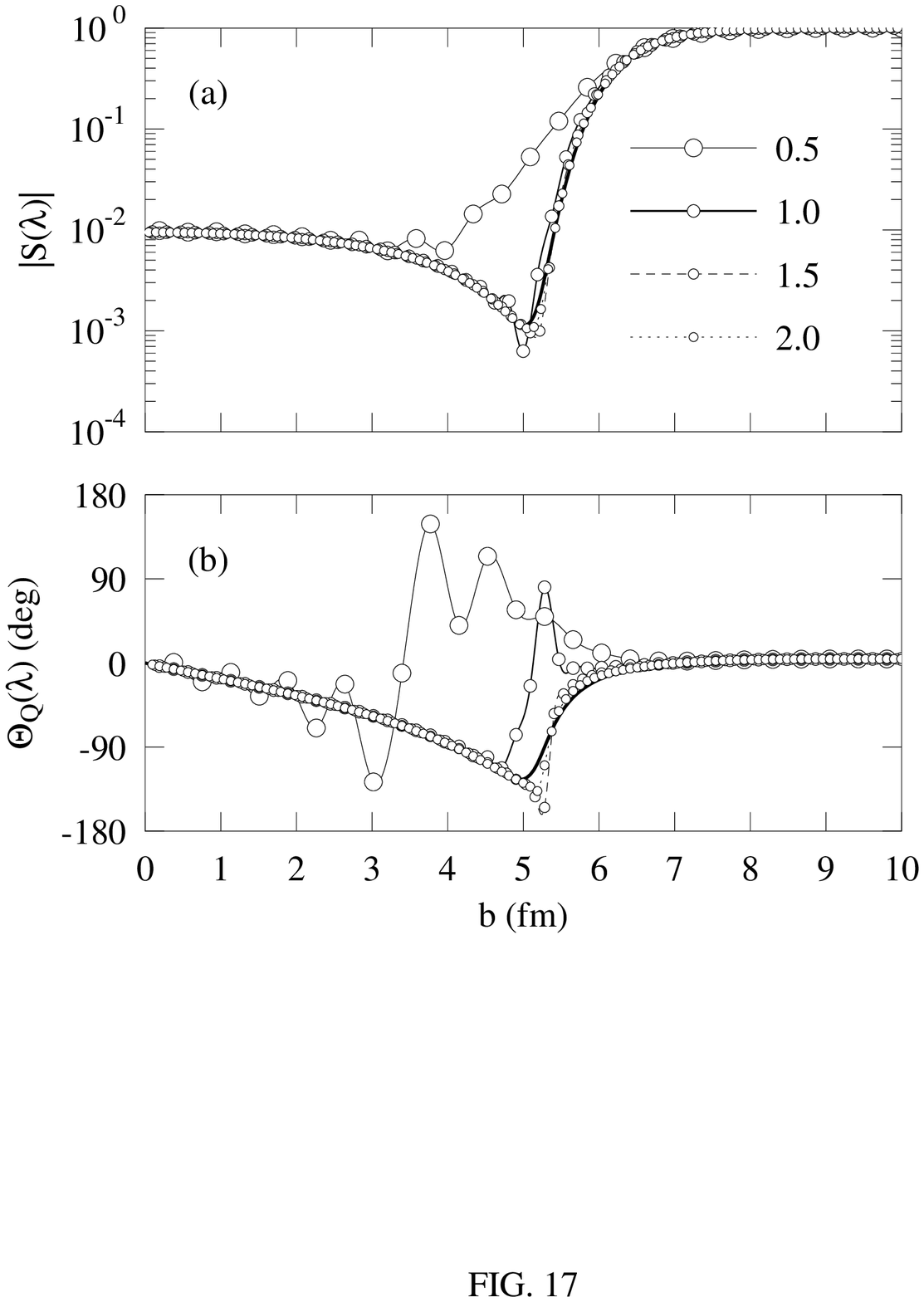, width=8.4cm, clip=}
\caption{The same as Fig. 10 for the complex optical potential at
$E_{\rm Lab}=200$ MeV.}
\end{figure}

\begin{figure} 
\label{FIG18}
\hspace*{-3mm}
\epsfig{file=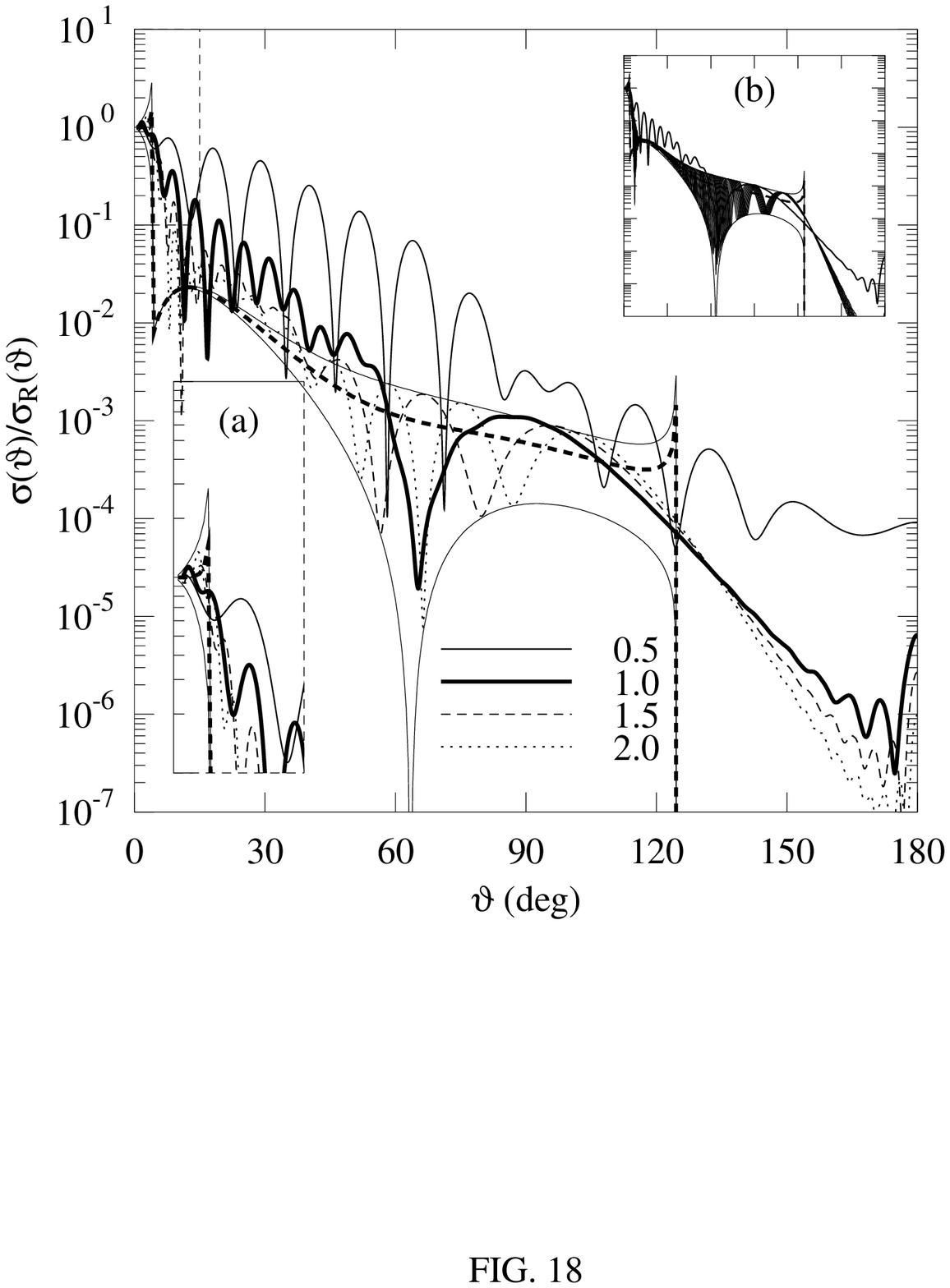, width=8.4cm, clip=}
\caption{The same as Fig. 5 for the full cross sections of the complex
optical potential at $E_{\rm Lab}=200$ MeV.}
\end{figure}

In the inset (b) of Figs. 18 and 20, one again observes that the
quantum calculations with $f$  values ranging from 3.0 to 4.0, very
well define the classical interference region, apart from problems
connected with the quantum illumination of the classical shadow regions.
Also in this case, the good definition of the interference region
allows one the test the classical origin of the different amplitudes
contributing to the quantum cross section by using only the
calculations of a standard optical potential code.

In the forward hemisphere, the values of the true far-side cross section
largely violates the classical interference limits.
This confirms the inappropriateness of using the rainbow terminology
for the interference patterns appearing in this and in the full cross
sections.

\begin{figure} 
\label{FIG19}
\hspace*{-3mm}
\epsfig{file=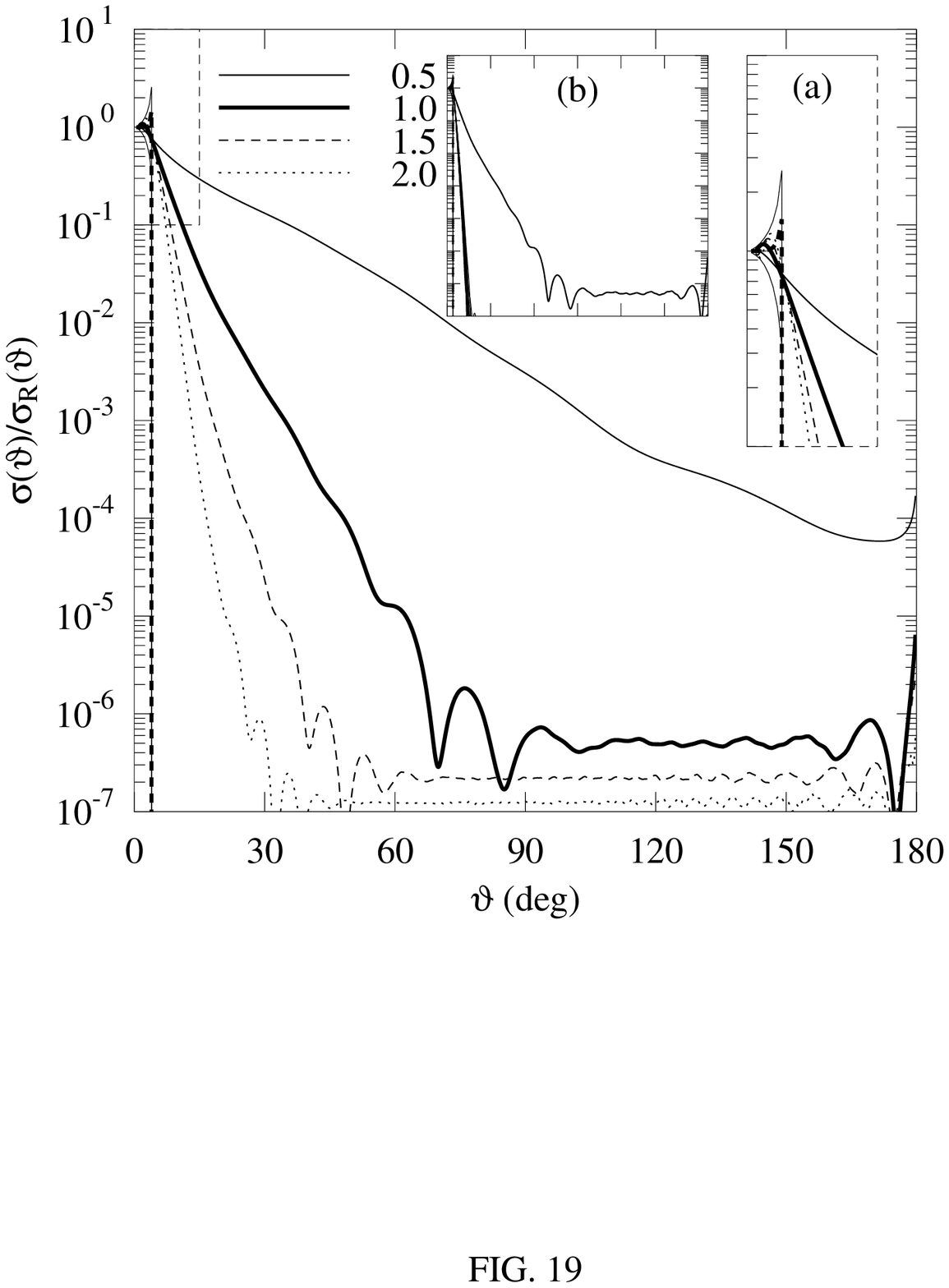, width=8.4cm, clip=}
\caption{The same as Fig. 5 for the near-side cross sections of the
complex optical potential at $E_{\rm Lab}=200$ MeV.}
\end{figure}

\begin{figure} 
\label{FIG20}
\hspace*{-3mm}
\epsfig{file=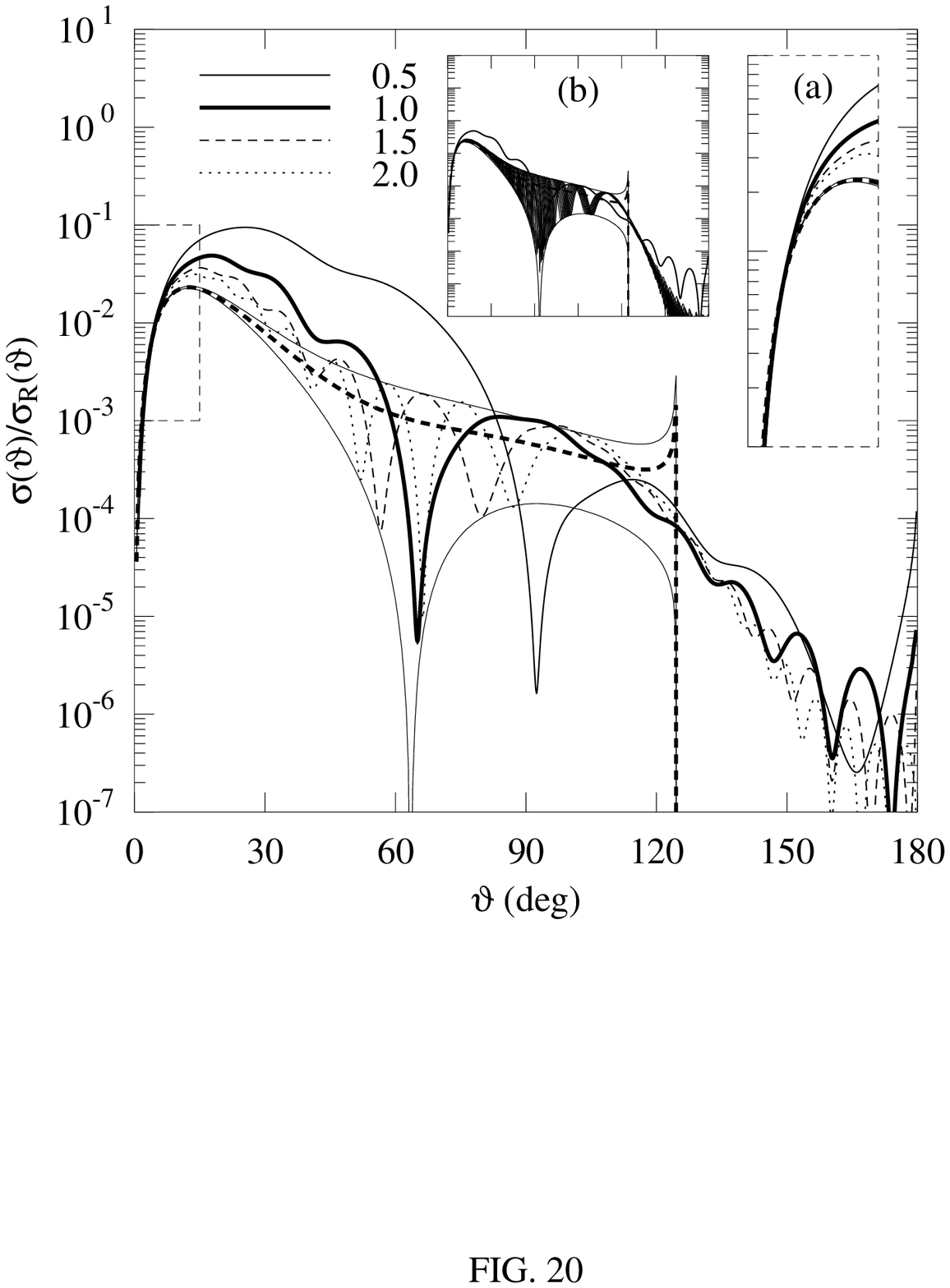, width=8.4cm, clip=}
\caption{The same as Fig. 5 for the far-side cross sections of the
complex optical potential at $E_{\rm Lab}=200$ MeV.}
\end{figure}

Figure 19 shows that, as in the previous case, the behavior of the
near-side cross section in the classical shadow region is largely
responsible for the violation of the interference limits of the full
cross section.

In the inset (b) of the same figure one can observe that, for $f$
values from 3.0 to 4.0 and for $\theta > \theta_C$, the almost constant
contribution to the near-side cross section is outside the plotted
area. 
Only the rapidly decreasing exponential contribution appears in a
restricted angular range above $\theta_C$.

\section{Conclusions}

The simple recipe of shrinking $\hbar$, in a conventional optical
potential calculation, provides useful information on the nature of the
different amplitudes contributing to the cross section.

By decreasing $\hbar$ the different characteristics of the cross
section smoothly change, with different rapidity.
In the major part of the angular interval, below some $\hbar$ value,
no further changes are observed in the cross sections with decreasing 
$\hbar$, apart from the sliding of the interference pattern within
well defined regions, with an increasing number of oscillations.
These are the characteristics connected with the realization of the
transition from the dynamical regime governed by the quantum mechanics
rules to that governed by classical mechanics ones.

The recipe can be easily implemented in any optical potential code,
providing a practical tool for a rapid check of the classical
properties of the cross section of a given potential.

The possibility of producing optical potential cross sections,
attributing different values to $\hbar$, can also be used as a
laboratory which provides useful cross sections for testing the
effectiveness of the semiclassical techniques currently used.

As one example of the tests that could be performed let us consider the
132 MeV case.
Following the Brink and Takigawa approximation, in this case the
oscillations appearing in the far side cross section arise from the
interference between the far-side contributions to the barrier and the
internal amplitudes\cite{ANN01}.
A similar result has also been obtained\cite{MIC00,MIC01} (with an
approximate calculation\cite{ALB82} of the barrier and internal
amplitudes) for the same and for several other optical potentials used
to describe the elastic scattering of light heavy-ions.
The results obatained in all these cases show that the barrier far-side
contribution is responsible for the Fraunhofer-like pattern appearing
in the barrier cross section, and this strongly suggests that it must
be regarded as a diffractive contribution.

The present analysis shows that by decreasing the value of $\hbar$ this
contribution smoothly changes, until it assumes a form which must be
identified with the contribution from classical-like far-side
trajectories.

In the Brink and Takigawa approximation the contribution from these
trajectories should be contained in the internal term, and this implies
that a critical $\hbar$ range exists in which the contribution migrates
from the barrier to the internal term. This migration is probably
connected with the change of the characteristics of the trajectories
described in the complex $r$ plane by the turning points of the radial
equation.

In this critical region, detailed semiclassical analyses of the cross
sections could provide interesting information on the validity, or on
the limits, of the semiclassical techniques, and allow us to achieve a
better understanding of the transition between the quantum and the
classical regimes.

\end{multicols}
\end{document}